\documentclass[11pt,notitlepage,nofootinbib]{revtex4-2}

\usepackage[dvipdf,dvips,dvipdfmx]{graphicx}
\usepackage{soul,xcolor}
	\definecolor{skyblue}{rgb}{0.0,0.74,1.0}
	\definecolor{hotpink}{rgb}{1.0,0.08,0.57}
\usepackage[breaklinks,linktoc=all,colorlinks=true,linkcolor=blue,citecolor=blue,urlcolor=blue]{hyperref}
\usepackage[acronym,toc]{glossaries}

\usepackage{xurl}
\usepackage{epsfig}
\usepackage{epstopdf}
\usepackage{float}
\usepackage{slashed}
\usepackage{cancel}
\usepackage{textcomp}
\usepackage{adjustbox}
\usepackage{enumitem}
\usepackage{bbm}
\usepackage{bbold}
\usepackage{indentfirst}
\usepackage{wrapfig}
\usepackage{comment}
\usepackage{pgfplots}
	\pgfplotsset{width=5cm,compat=1.9}
\usepackage{bm}
\usepackage{fancyhdr}
\usepackage{amsmath}
\usepackage{amssymb}
\usepackage{calc}
\usepackage{mathtools}
\usepackage{gauss}
\usepackage{physics}
\usepackage{gensymb}
\usepackage{esint}
\usepackage{tikz-feynman}
\usepackage{tensor}
\usepackage{xfrac}
\usepackage{kbordermatrix}
	\renewcommand{\kbldelim}{(}
	\renewcommand{\kbrdelim}{)}

\newcommand{\nocontentsline}[3]{}
\newcommand{\tocless}[2]{\bgroup\let\addcontentsline=\nocontentsline#1{#2}\egroup}

\newcommand{\nnr}{\nonumber\\}

\newcommand{\Array}[1]{\begin{pmatrix} #1 \end{pmatrix}}
\newcommand{\trans}{\enskip\rightarrow\enskip}

\newcommand{\eqspace}{\qquad\qquad\qquad}
\newcommand{\sml}[1]{\mbox{$#1$}}

\def\lsim{\mathrel{\mathpalette\@versim<}}
\def\gsim{\mathrel{\mathpalette\@versim>}}

\newcommand{\sumoverset}[2]{\overset{\scriptstyle #1\mathstrut}{#2}}

\newcommand{\Com}[2]{\left[#1,#2\right]}
\newcommand{\tCom}[2]{\big[#1,#2\big]}

\newcommand{\tACom}[2]{\big\{#1,#2\big\}}

\newcommand{\dx}{\dd^4 x\,}
\newcommand{\bp}{{\bm p}}
\newcommand{\bq}{{\bm q}}
\newcommand{\hc}{\text{h.c.}}
\newcommand{\Mp}{M_\text{P}}
\newcommand{\mmet}{\eta_{\alpha\beta}}

\newcommand{\met}{g_{\alpha\beta}}

\newcommand{\gdet}{\sqrt{-g}}
\newcommand{\hab}{h_{\alpha\beta}}

\newcommand{\hs}[1]{h_{#1}{}^{#1}}
\newcommand{\Hab}{H_{\alpha\beta}}
\newcommand{\iHab}{H^{\alpha\beta}}
\newcommand{\Hs}[1]{H_{#1}{}^{#1}}
\newcommand{\yab}{\psi_{\alpha\beta}}
\newcommand{\iyab}{\psi^{\alpha\beta}}

\newcommand{\Ord}{\mathcal{O}}

\newcommand{\Lag}{\mathcal{L}}
\newcommand{\Act}{\mathcal{S}}
\newcommand{\Ham}{\mathcal{H}}
\newcommand{\Smat}{S}
\newcommand{\Tmat}{T}
\newcommand{\FS}{\mathcal{V}}
\newcommand{\FSp}{\mathcal{V}_{\text{phys}}}
\newcommand{\bFSp}{\bar{\mathcal{V}}_{\text{phys}}}
\newcommand{\BChg}{\mathcal{Q}}

\newcommand{\Par}{P}
\newcommand{\Tim}{T}
\newcommand{\PT}{\Par\Tim}

\newcommand{\IPo}{\eta}
\newcommand{\hIPo}{\rho}
\newcommand{\eIPo}{\sigma}
\newcommand{\psIP}{\tensor[_\IPo]{\langle\cdot\rangle}{}}

\newcommand{\psc}{{\mathchoice
	{\includegraphics[height=2.0ex]{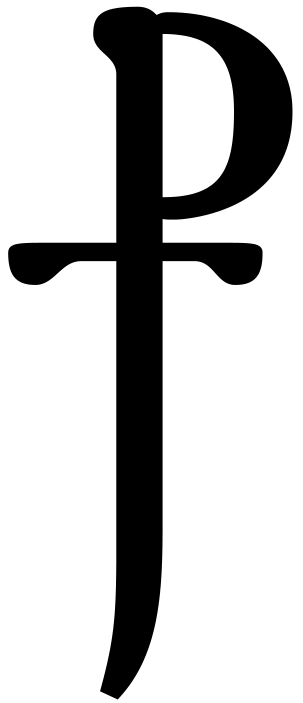}}
	{\includegraphics[height=2.0ex]{taurho.PNG}}
	{\includegraphics[height=1.4ex]{taurho.PNG}}
	{\includegraphics[height=1.0ex]{taurho.PNG}}}}
\newcommand{\psbra}[1]{\tensor[_{\IPo}]{\bra{#1}}{}}
\newcommand{\psbraket}[2]{\tensor[_{\IPo}]{\braket{#1}{#2}}{}}
\newcommand{\psmel}[3]{\tensor[_{\IPo\!}]{\mel{#1}{#2}{#3}}{}}

\begin{document}

\title{Unitarity through $\PT$ symmetry in\\Quantum Quadratic Gravity}

\author{Jeffrey \surname{Kuntz}}
\email{jkuntz@mpi-hd.mpg.de}
\affiliation{Max-Planck-Institut f\"ur Kernphysik (MPIK), Saupfercheckweg 1, 69117 Heidelberg, Germany}

\date{\today}

\begin{abstract}
Theories described by non-Hermitian Hamiltonians are known to possess strictly positive energy eigenvalues and exhibit unitary time evolution if the Hamiltonian is symmetric under discrete parity and time ($\PT$) transformation. In this work, we demonstrate how quantum quadratic gravity, a theory that generally violates unitarity when viewed as a Hermitian quantum field theory, can be complex-deformed into such a $\PT$-symmetric theory with an action that consists of a ghost-less Hermitian free part and non-Hermitian interactions. Paying special attention to the gauge symmetry present in the theory, we quantize in the covariant operator formalism after suggesting how the framework might be extended to the pseudo-Hermitian picture. We find compelling evidence that the resulting quantum theory possesses a unitary inner product and a sensible interpretation of quantum probability, thus avoiding the ghost problem present in the Hermitian formulation of quadratic gravity.
\end{abstract}


\maketitle

{\hypersetup{linkcolor=black}\tableofcontents}

\clearpage

\section{Introduction}

It is well-known that, despite its overwhelming success as a classical theory, Einstein's General Relativity (GR) does not represent the basis for a satisfactory theory of quantum gravity due to the simple fact that it is power-counting non-renormalizable. It has however been shown that supplementing the classical Einstein-Hilbert action, which is linear in powers of the Ricci scalar, with all independent curvature invariants that are quadratic in curvature tensors leads to a (perturbatively) renormalizable theory \cite{Stelle1977}, referred to here as quadratic gravity (QG). This remarkable fact suggests that QG is a superior description of gravity, as it very generally reproduces or even improves on all of the well-established low-energy predictions of GR \cite{Salvio2014,Alvarez-Gaume2016,Salvio2018,Salvio2018b,Salvio2022,Menezes2023}, while also putting our understanding of gravity on par with the Standard Model (SM) forces in terms of its renormalizability i.e.\ predictivity at the quantum level. 

There is however more to the story -- the same higher powers of curvature that render QG renormalizable contain four derivatives acting on the metric and thus generally exhibit the so-called Ostrogradsky instability at the classical level \cite{Ostrogradsky1850,Woodard2015}. This implies that the Hilbert space of the associated quantum theory contains ghost states with negative norm. Though it is often claimed that this leads to a breakdown of unitarity, the more precise statement is that the presence of ghost states makes it impossible to interpret quantum probability in the usual way. This ``ghost problem'' actually disappears when one considers QG an effective theory that is only valid at energies below the mass of the ghost \cite{Kuntz2022b,Kubo2023,Kubo2024}, however, it is probably impossible to reconcile negative norms at arbitrary scales within the standard quantum field theory (QFT) framework. Some of the original works that address the ghost problem include the models of Lee and Wick \cite{Lee1969a,Lee1970}, and Boulware, Horowitz, and Strominger \cite{Boulware1983}, while more recent notable resolutions have been proposed by Donoghue and Menezes \cite{Donoghue2018b,Donoghue2019a,Donoghue2019,Donoghue2021}, Anselmi \cite{Anselmi2017,Anselmi2018,Anselmi2018a,Anselmi2022}, and most importantly for this work, by Bender, Mannheim, and collaborators \cite{Bender1998,Bender1999,Bender2004,Bender2005,Bender2007,Bender2019,Mannheim2012,Mannheim2018,Mannheim2020a,Mannheim2023,Mannheim2023a,Kleefeld2023} (solutions in this spirit are also considered by Salvio and Strumia \cite{Salvio2016,Salvio2019a}).

It has been demonstrated that the standard requirement of Hermiticity for a Hamiltonian, $\Ham^\dag=\Ham$, is a sufficient though not necessary condition for guaranteeing a set of strictly positive eigenvalues and unitary time evolution \cite{Bender1998,Bender1999}. A more general axiom for ensuring these physical requirements in a QFT is that the theory's Hamiltonian must possess an unbroken anti-linear symmetry. Interestingly, if one assumes complex Lorentz invariance and that the inner product on a given Hilbert space conserves probability, then such an anti-linear symmetry emerges naturally, namely, symmetry under combined charge, parity, and time ($C\PT$) transformations \cite{Bender2007,Bender2019,Mannheim2018b}. Naturally, this is equivalent to $\PT$ symmetry for trivially $C$-symmetric (neutral) theories such as QG and all other theories that we will consider in this work. It is important to note that Hermitian operators are necessarily $\PT$-symmetric and that assuming a $\PT$-symmetric paradigm does not invalidate any of the usual Hermitian QFT, rather, it extends the space of physically viable theories (those with strictly positive Hamiltonian eigenvalues and a positive-definite Hilbert space). This picture can also extend the range of validity of established Hermitian theories when their renormalized couplings run into a non-Hermitian but $\PT$-symmetric phase (see e.g.\ the Lee model \cite{Bender2005a} or the unstable effective potentials in the SM Higgs and supergravitational gravitino sectors \cite{Bender2016}). Indeed, $\PT$ symmetry and renormalization are generally complimentary in quantum field theory \cite{Bender2018b,Bender2021}, though we will not discuss this relationship in detail here. The purpose of this work is demonstrate how QG might be reinterpreted as a $\PT$-symmetric QFT wherein a sensible interpretation of probability at the quantum level can be established, thus rendering QG a healthy QFT under this paradigm. 

In Sec.\ \ref{sec:GP} we will examine how the ghost problem manifests in a fourth-order toy model and see how it may be avoided through a complex deformation of the original Hermitian theory. Before moving on to QG, in Sec.\ \ref{sec:PTSym} we will review the $\PT$-symmetric/pseudo-Hermitian framework that will be required later and discuss how the methods of covariant quantization in the Becchi-Rouet-Stora-Tyutin (BRST) formalism must be modified in that framework. This will prepare us for the quantization and identification of the transverse physical subspace of QG in Sec.\ \ref{sec:QG}. Sec.\ \ref{sec:UniInt} is then dedicated to the construction of the modified inner product and a demonstration of perturbative unitarity in a truncated version of the interacting non-Hermitian theory of quadratic gravity that we previously established.

\section{The Ghost Problem} \label{sec:GP}

\subsection{The Pais-Uhlenbeck oscillator} \label{subsec:PUo}

We begin our discussion with a look at a classic example of a theory that exhibits the Ostrogradsky instability -- the Pais-Uhlenbeck oscillator \cite{Pais1950}. This theory is described by the following fourth-order (in time derivatives) action of a dynamical coordinate $z(t)$ subject to an arbitrary potential $V(z)$
\begin{align} \label{APUz}
\Act_\text{PU} = \int\dd t\left[\frac{1}{2(\omega_1^2 - \omega_2^2)}\Big(\ddot{z}^2 - (\omega_1^2 + \omega_2^2)\dot{z}^2 + \omega_1^2\omega_2^2\,z^2\Big) - V(z)\right] \,,
\end{align}
where $\omega_1$ and $\omega_2$ are constant angular frequencies (masses). Due to the presence of higher derivatives, this action actually possesses more than one independent degree of freedom, so to get a better feeling for the physics at play, it is convenient to re-express it as a second-order theory in more independent variables. We forego the often used Ostrogradsky procedure that jumps straight to the Hamiltonian formulation and instead first rewrite the action above in the equivalent form
\begin{align} \label{APUaux}
\Act_\text{PUaux} = \int\dd t\left[u\ddot{z} + \omega_1^2\,uz + \frac12\Big(\dot{z}^2 - \omega_1^2\,z^2 - (\omega_1^2 - \omega_2^2)u^2\Big) - V(z)\right] \,,
\end{align}
where we have introduced an auxiliary coordinate $u(t)$ that, when integrated out using the equation of motion
\begin{align} \label{uEOM}
u = \frac{1}{\omega_1^2 - \omega_2^2}\big(\ddot{z} + \omega_1^2\,z\big) \,,
\end{align}
returns the original action \eqref{APUz}. We may also put this action in a more canonical form by redefining $z(t)$ as
\begin{align} \label{zdec}
z(t) = x(t) + u(t) \,,
\end{align}
which diagonalizes \eqref{APUaux} and leaves us with
\begin{align} \label{APUxu}
\Act_\text{PUdiag} = \int\dd t\left[\frac12\Big(\dot{x}^2 - \omega_1^2\,x^2\Big) - \frac12\Big(\dot{u}^2 - \omega_2^2\,u^2\Big) - V(x + u)\right] \,,
\end{align}
after integrating by parts and assuming that surface terms may be neglected. This action clearly describes the dynamics of two independent coordinates $x$ and $u$, the latter of which may be identified as an Ostrogradsky ghost by the relative minus sign carried by its kinetic term.

The equations of motion (EOMs) obtained from \eqref{APUxu},
\begin{align} \label{xuEOMs}
\ddot{x} = -\big(\partial_xV + \omega_1^2\,x\big) \eqspace \ddot{u} = \partial_uV - \omega_2^2\,u \,,
\end{align}
allow us to see how the Ostrogradsky instability manifests. Given the dependence of the potential, we find that $\partial_xV=\partial_uV$ and thus what is a restoring force for $x$ is an anti-restoring force for $u$. When $\partial_uV>\omega_2^2\,u$, this implies the existence of runaway solutions that result from a classically negative kinetic energy associated with $u$ and the fact that energy is conserved in the system.

To see how this instability manifests at the quantum level, we switch to the Hamiltonian framework and focus on the free part of the theory, where the canonical momenta are defined by
\begin{align}
p_x(t) = \dot{x}(t) \eqspace p_u(t) = -\dot{u}(t)
\end{align}
and the Hamiltonian is simply
\begin{align} \label{HPUxu}
\Ham_0 = \int\dd t\,\frac12\Big(p_x^2 + \omega_1^2\,x^2 - p_u^2 - \omega_2^2\,u^2\Big) \,.
\end{align}
We may then quantize the theory by imposing the canonical commutation relations
\begin{align} \label{comxu}
&\Com{x}{p_x} = i \eqspace \Com{u}{p_u} = i \,.
\end{align}

Before proceeding, it is worth commenting on our choice of canonical variables that appear in the commutators above. Despite the fact that our auxiliary action \eqref{APUaux} describes the same physical system as the original fourth order action (it leads to an equivalent set of EOMs), the way in which we arrive at our second order Hamiltonian \eqref{HPUxu}, which also appears in the original Pais-Uhlenbeck study \cite{Pais1950}, differs slightly from the standard Ostrogradsky treatment of fourth-order theories \cite{Ostrogradsky1850,Woodard2015}. In this method, one establishes a second-order Hamiltonian directly from the fourth-order action, noting that the four pieces of initial data required to solve the fourth-order EOM imply that there must be four canonical variables. For a general, non-degenerate Lagrangian $\Lag(z,\dot{z},\ddot{z})$, these are defined by:
\begin{align}
&X_1(t) = z(t)& &P_1(t) = \frac{\partial\Lag}{\partial\dot{z}(t)} - \frac{\dd}{\dd t}\frac{\partial\Lag}{\partial\ddot{z}(t)}& &X_2(t) = \dot{z}(t)& &P_2(t) = \frac{\partial\Lag}{\partial\ddot{z}(t)} \,.
\end{align}
In the particular case of the Pais-Uhlenbeck oscillator action \eqref{APUz}, written as $\Act_\text{PU} = \int\dd t\Lag_\text{PU}$, these definitions yield the off-diagonal second-order free Hamiltonian
\begin{align}
\Ham_0^{(\text{Ost})} &= \int\dd t\bigg(\sum_iP_i\dot{X}_i - \Lag_\text{PU,0}\bigg) \nnr
&= \int\dd t\bigg(P_1X_2 + \frac{\omega_1^2 - \omega_2^2}{2}P_2^2 - \frac{1}{2\big(\omega_1^2 - \omega_2^2\big)}\Big(\omega_2^2\omega_2^2X_1^2 - \big(\omega_1^2 + \omega_2^2\big)X_2^2\Big)\bigg) \,,
\end{align}
which has also been rigorously derived using Dirac's method of constraints in \cite{Mannheim2000,Mannheim2005a}. The pairs of canonical variables appearing in this Hamiltonian satisfy the required Poisson bracket relations, $\big\{X_i,P_j\big\}=\delta_{ij}$, and, crucially, are related to the variables $(x,p_x)$ and $(u,p_u)$ in \eqref{HPUxu} by a canonical transformation, as identified by Smilga in \cite{Smilga2009}. It is straightforward to confirm the relations $\{x,p_x\}=\{u,p_u\}=1$ using the inverse of Smilga's transformation, which identifies $(x,p_x)$ and $(u,p_u)$ as canonical variable pairs and thus validates our imposition of the commutation relations \eqref{comxu}.

For completeness, we may also construct the free Hamiltonian associated with the second-order auxiliary field action \eqref{APUaux},
\begin{align}
\Ham_0^{(\text{aux})} = -\int\dd t\bigg(\bar{p}_z\bar{p}_u + \frac12\bar{p}_u^2 - \frac{\omega_1^2}{2}\bar{z}^2 + \omega_1^2\bar{z}\bar{u} - \frac{\omega_1^2 - \omega_2^2}{2}\bar{u}^2\bigg) \,,
\end{align}
whose coordinates are defined as
\begin{align}
&\bar{z} = z& &\bar{p}_z = \dot{z} - \dot{u}& &\bar{u} = u& &\bar{p}_u = -\dot{z} \,.
\end{align}
We find that these fit into the expected canonical pairs $(\bar{z},\bar{p}_z)$ and $(\bar{u},\bar{p}_u)$, as they are related to the diagonal $x$--$u$ basis in \eqref{APUxu} via the canonical transformations
\begin{align}
&\bar{z} = x + u& &\bar{p}_z = p_x& &\bar{u} = u& &\bar{p}_u = -p_x + p_u \,.
\end{align}
We thus confirm that the auxiliary field procedure that we employ in this work yields a satisfactory basis of canonical coordinates just as the Ostrogradsky procedure does. Since our auxiliary $z$--$u$ basis, diagonal $x$--$u$ basis, and the Ostrogradsky/Dirac constraints $X_1$--$X_2$ basis are all related via canonical transformation, they all represent equally valid descriptions of the theory at hand. We employ the auxiliary field method because it allows us to diagonalize and obtain a clear description of theory's DOFs already at the level of the action, as opposed to the Hamiltonian-based Ostrogradsky interpretation of the theory.

Moving forward, we consider the Heisenberg picture, where time dependence is carried by operators that act on quantum states, and express our canonical variables as
\begin{align} \label{xudecomp}
\begin{aligned}
&x(t) = \frac{1}{\sqrt{2\omega_1}}\Big(a_x e^{-i\omega_1t} + a_x^\dag e^{i\omega_1t}\Big)\qquad& &u(t) = \frac{1}{\sqrt{2\omega_2}}\Big(a_u e^{-i\omega_2t} + a_u^\dag e^{i\omega_2t}\Big) \\
&p_x(t) = -i\sqrt{\frac{\omega_1}{2}}\Big(\,a_x e^{-i\omega_1t} - a_x^\dag e^{i\omega_1t}\Big)& &p_u(t) = i\sqrt{\frac{\omega_2}{2}}\Big(a_u e^{-i\omega_2t} - a_u^\dag e^{i\omega_2t}\Big) \,,
\end{aligned}
\end{align}
where $a_i^\dag$ and $a_i$ are time-independent creation and annihilation (c/a) operators. Inserting these decompositions into \eqref{HPUxu} yields the Heisenberg picture Hamiltonian operator
\begin{align}
\Ham_0 = \frac12\big(\omega_1\,a_x^\dag a_x - \omega_2\,a_u^\dag a_u\big) \,,
\end{align}
and consistency with the Heisenberg equation, $-i\dot{\phi}=\tCom{\Ham_0}{\phi}$ where $\phi=\{x,u\}$, yields the non-vanishing commutation relations between state operators,
\begin{align} \label{comaxu}
\Com{a_x}{a_x^\dag} = 1 \eqspace \Com{a_u}{a_u^\dag} = -1 \,.
\end{align}

The minus sign in the second commutation relation of \eqref{comaxu} confirms $u$'s status as a ghost and is what leads to the theoretical issues at the quantum level that we refer to as the ghost problem. To see where these issues occur, we first recall that one is in fact free to define either state operator as the creation or annihilation operator i.e.\ define the vacuum state in a theory based on either
\begin{align} \label{pmvacs}
a_j\ket{0}_+ = 0 \qquad\text{or}\qquad a_j^\dag\ket{0}_- = 0 \,.
\end{align}
This choice has important consequences as it defines the sign of the norm and the energy eigenvalues of a system. For the healthy particle ($a_x$), one naturally chooses the first ($+$) option as it corresponds to a positive norm and positive eigenvalues, while for the ghost the choice is at first not so clear since the $\ket{0}_+$ ($\ket{0}_-$) option corresponds to a negative (positive) norm and positive (negative) energy eigenvalues:
\begin{align}
\tensor[_\pm]{\braket{a_u}{a_u}}{_\pm} = \mp 1 \eqspace \Ham_0\tensor{\ket{a_u}}{_\pm} = \pm \omega_2 \,.
\end{align}

There is however some criteria that clears up this ambiguity for the ghost when it cohabitates a theory with healthy particles. If we write down the propagators for each type of particle,
\begin{align}
&\tensor[_\pm]{\mel{0}{\text{T}\,a_x(t)a_x(t')}{0}}{_\pm} = \frac{i}{2}\int\dd E\frac{e^{-iE(t - t')}}{E^2 - \omega_1^2 \pm i\epsilon} \\
&\tensor[_\pm]{\mel{0}{\text{T}\,a_u(t)a_u(t')}{0}}{_\pm} = -\frac{i}{2}\int\dd E\frac{e^{-iE(t - t')}}{E^2 - \omega_2^2 \pm i\epsilon} \,,
\end{align}
we see that they differ only in the leading minus sign and that they share the same pole structure; a pole located in the upper (lower) half complex plane for the $\ket{0}_+$ ($\ket{0}_-$) vacuum. It is thus necessary to select the same vacuum for each type of particle if we wish to maintain a consistent $i\epsilon$ prescription for Feynman diagrams that contain both species \cite{Stelle1977,Salvio2018,Zimmermann1968}. After requiring positive energy eigenvalues and norms for the healthy particle, we are thus left with positive energy eigenvalues and negative norms for the ghost, despite the fact that this breaks correspondence with the classical system where negative kinetic energy is associated with the ghost.

Naturally, the ghost problem also manifests in the Schrödinger picture where the states of our system are governed by a quantum wavefunction $\psi$ that is subject to the Schrödinger equation, $\Ham_0\psi_n = \lambda_n\psi_n$, where the Hamiltonian operator is defined by replacing $p_x\to-i\partial_x$ and $p_u\to-i\partial_u$ in \eqref{HPUxu}.
The normalized $n$-th excited state wavefunctions and energy eigenvalues of this Hamiltonian are found to be
\begin{align}
&\psi_n(x,u) = \frac{(\omega_1\omega_2)^{1/4}}{2^n\,n!\sqrt{\pi}}H_n(\sqrt{\omega_1}x)H_n(-i\sqrt{\omega_2}u)e^{-\frac12\left(\omega_1x^2 - \omega_2u^2\right)} \qquad \lambda_n = (n + 1/2)(\omega_1 + \omega_2) \,,
\end{align}
where $H_n$ is the Hermite polynomial of order $n$. At this stage, we can already spot a serious issue in that these wavefunctions are non-normalizable with respect to the basic definition involving integration over the real axis, $\braket{\psi_m}{\psi_n}=\int_\mathbb{R}\dd x\dd u(\psi_m^*\psi_n)$, as they diverge in the $u\to\pm\infty$ limit. This fact might also seem to imply serious problems involving the vacuum itself since, as pointed out in \cite{Mannheim2023a}, the integral that determines the vacuum normalization in QFT, $\braket{0}{0}=1$, is linked to the normalization of the ground state wave function. However, it has been known for some time how to deal with this situation, as it also arises in second-order gauge theories that contain unphysical states with positive Gaussian signature in their complete Fock space. As established by Arisue et.\ al.\ \cite{Arisue1987}, the normalization issue related to wavefunctions with positive Gaussian signature can be avoided by integrating the associated variable over the imaginary axis\footnote{We thank Taichiro Kugo for pointing out this fact.}:
\begin{align} \label{psinorm}
\braket{\psi_m}{\psi_n} = \int_\mathbb{R}\dd x\int_\mathbb{I}\dd u \big(\psi_m^*\psi_n\big) = (-1)^m\delta_{nm} \,.
\end{align}
With this, we resolve any vacuum-related issues, but we also encounter an indefinite inner product metric, a feature that also manifests in the Heisenberg picture due to the norm $\braket{a_u}{a_u}=-1$. This means that is impossible to define quantum probability with the standard Born rule,
\begin{align} \label{Born}
P_\psi = \frac{\big|\braket{\psi}{\chi}\big|^2}{\braket{\chi}{\chi}} \,.
\end{align}
Here, $P_\psi$ represents the probability of finding the state $\ket{\chi}$ in the particular configuration $\ket{\psi}$, which becomes meaningless if $\braket{\chi}{\chi}\leq0$. This breakdown of the probability interpretation at the quantum level is what truly constitutes the ghost problem.

\subsection{Complex deformation} \label{subsec:compdef}

In this work, we propose a solution to the ghost problem that follows the work of Bender and Mannheim \cite{Bender2008}, revolving around the crucial assumption that \textit{the ghost is actually anti-Hermitian}. This complex deformation of the original theory is implemented by defining new Hermitian variables $y(t)$ and $p_y(t)$ that are related to the original ghostly variables through a similarity transformation using the operator $e^{-R}$ with $R=\!-\frac{\pi}{2}p_uu$:
\begin{align} \label{uytrans}
y(t) \equiv e^{-R}\,u(t)\,e^{R} = -iu(t) \eqspace p_y(t) \equiv e^{-R}\,p_u(t)\,e^{R} = ip_u(t) \,.
\end{align}
It is important that our field redefinition is based on a  similarity transformation as this preserves the structure of the canonical commutation relations. Indeed, applying the transformation to \eqref{comxu}, we find that the $x$ commutator is unaffected while the $u$ commutator becomes $\Com{y}{p_y} = i$. This identification of $u(t)$ as an anti-Hermitian field also has the important implication that, per \eqref{zdec}, the original variable $z(t)$ that appears in the fourth-order theory is necessarily complex. Indeed, another way to view the rewriting of the original action is that we have actually assumed that $z(t)$ is complex and split it into its independent real and complex parts, since with \eqref{zdec} and \eqref{uytrans}, we have that $\text{Re}[z]=x$ and $\text{Im}[z]=y$.

After transforming the ground state wavefunction,
\begin{align}
\tilde{\psi}_n(x,y) \equiv e^{-R}\,\psi_n(x,u)\,e^{R} = \frac{(\omega_1\omega_2)^{1/4}}{2^n\,n!\sqrt{\pi}}H_n(\sqrt{\omega_1}x)H_n(\sqrt{\omega_2}y)e^{-\frac12\left(\omega_1x^2 + \omega_2y^2\right)} \,,
\end{align}
we find the desired positive-definite inner product metric after integrating purely over the real axis in the standard way:
\begin{align}
\big\langle\tilde{\psi}_m\big|\tilde{\psi}_n\big\rangle = \int_\mathbb{R}\dd x\dd y \big(\tilde{\psi}_m^*\tilde{\psi}_n\big) = \delta_{nm} \,.
\end{align}
The transformed free Hamiltonian also appears in the healthy ghost-free form
\begin{align} \label{HPUxy}
\tilde{\Ham}_0 \equiv e^{-R}\,\Ham_0\,e^{R} = \int\dd t\bigg[\frac12\Big(p_x^2 + \omega_1^2\,x^2 + p_u^2 + \omega_2^2\,u^2\Big)\bigg]
\end{align}
and maintains the same eigenvalues as the original un-rotated version:
\begin{align}
\tilde{\Ham}_0\,\tilde{\psi}_n = (n + 1/2)(\omega_1 + \omega_2)\tilde{\psi}_n \,.
\end{align}

Naturally, this all applies in the Heisenberg picture version of the theory as well, where the identification of the ghost variables as anti-Hermitian implies an inconsistency in the decomposition \eqref{xudecomp} since the c/a operators can no longer be Hermitian conjugates of each other. This inconsistency is readily resolved by first rotating to $y$ and $p_y$, then decomposing these Hermitian variables in the standard way as
\begin{align} \label{xydecomp}
y(t) = \frac{1}{\sqrt{2\omega_2}}\Big(a_y e^{-i\omega_2t} + a_y^\dag e^{i\omega_2t}\Big) \eqspace p_y(t) = -i\sqrt{\frac{\omega_2}{2}}\Big(\,a_y e^{-i\omega_2t} - a_y^\dag e^{i\omega_2t}\Big) \,,
\end{align}
where $a_y$ and $a_y^\dag$ are indeed Hermitian conjugates. With this, one can easily show that the free $y$ behaves as a healthy particle with a positive commutation relation $\tCom{a_y}{a_y^\dag}=1$ and positive eigenvalue $\tilde{\Ham}_0\ket{a_y}_+=\omega_2\ket{a_y}_+$.

Thus, after redefining our ghost variables per the similarity transformation \eqref{uytrans}, the free Hamiltonian \eqref{HPUxy} corresponds to two standard healthy harmonic oscillators and all of the pathologies associated with the ghost problem that arise from the troublesome minus sign are absent. This disappearance of the ghost problem in the free theory comes as a direct result of our rewriting the fourth-order theory using the auxiliary field formulation and diagonalizing at the level of the action in \eqref{APUxu}. When the Ostrogradsky procedure is used instead, as it has been in previous studies of the $\PT$-symmetric Pais-Uhlenbeck oscillator, the free part of the theory appears in a different (though dynamically equivalent) form with factors of $i$. As we will see in later sections, the present formulation with a Hermitian free part makes the demonstration of unitarity far more transparent than it would be in the Ostrogradsky formulation. Of course, removing ghosts from the free part of the theory is not the end of the story however. The complex deformation leads to factors of $i$ in the potential of \eqref{APUxu},
\begin{align}
e^{-R}V(x + u)e^{R} = V(x + iy) \,,
\end{align}
which leaves us with a non-Hermitian interacting Hamiltonian and serious concerns about the complex nature of inner products, and thus also unitarity. 

\section{Non-Hermitian Theories with $\PT$ Symmetry} \label{sec:PTSym}

\subsection{General considerations} \label{subsec:gencons}

It may seem that there is nothing to be gained from reinterpreting a ghostly theory as non-Hermitian theory since we must trade the ghost problem for new issues that violate the core tenets of commonly accepted quantum theory. However, a resolution to these new concerns can arise from the fact that, though Hermiticity is traditionally regarded as a fundamental requirement for a physical Hamiltonian to exhibit unitary time evolution, Hermitian theories are in fact just a subset of a broader class of theories that also satisfy this crucial property. Assuming that the Hamiltonian is symmetric ($\Ham^T=\Ham$) and that it admits a complete basis of bi-orthonormal eigenstates $\{\ket{\psi_n},\ket{\psi'_n}\}$:
\begin{align} \label{biortho}
\Ham\ket{\psi_n} = \lambda_n\ket{\psi_n} \qquad\text{and}\qquad \Ham^\dag\ket{\psi'_n} = \lambda_n^*\ket{\psi'_n} \qquad\text{with}\qquad \braket{\psi'_m}{\psi_n} = \delta_{nm} \,,
\end{align}
the most general constraint is actually that the Hamiltonian must respect some anti-linear symmetry i.e.\ $\tCom{\Ham}{A}=0$ for some anti-linear symmetry operator $A$. Of course, physicality also relies on the Hamiltonian possessing strictly real and positive eigenvalues, $\lambda_n^*=\lambda_n$. This requirement is satisfied if the eigenstates of $\Ham$ are also eigenstates of $A$, in which case we say that the symmetry of $A$ is unbroken. However, this requirement is sometimes difficult to demonstrate exactly and must be verified numerically, since it requires knowledge of the complete basis of Hamiltonian eigenstates \cite{Bender2005}.

$\PT$ transformation is a particularly appealing candidate for such a fundamental anti-linear symmetry as it emerges naturally when one extends the real homogeneous Lorentz group to the complex domain \cite{Bender2019}. $\Par$ stands for discrete parity reflection and reverses the sign of the spatial components of a four-vector, while $\Tim$ represents discrete time reflection and reverses the sign of temporal four-vector components (as well as the imaginary unit $i$). These properties are summarized by
\begin{align}
\begin{array}{c|ccc}
& t & {\bm x} & i \\
\hline
\Par & + & - & + \\
\Tim & - & + & -
\end{array}_{\;{\textstyle .}}
\end{align}
Though combined $\PT$ symmetry is aesthetically appealing as fundamental symmetry of nature since it carries a very natural interpretation in terms of physical principles as opposed to the more abstract mathematical notion of Hermiticity, its most motivating feature is perhaps the simple fact that $\PT$-symmetric non-Hermitian dynamics have already been observed in real-world physical experiments (see e.g.\ \cite{Guo2009,Ruter2010,Bittner2012,Doppler2016,Xu2016}). Naturally, this experimental evidence strongly hints that the same principle may be applied to resolving issues in high energy physics, in particular the ghost problem, as well.

To apply the notions of positivity and unitarity through anti-linear $\PT$ symmetry, we first consider the standard inner product between the Schrödinger equation and its Dirac-adjoint with a non-Hermitian Hamiltonian,
\begin{align} \label{badHE}
\partial_t\!\braket{\psi_n}{\psi_m} = -i\mel{\psi_n}{\big(\Ham - \Ham^\dag\big)}{\psi_m} \,.
\end{align}
It is crucial that we retain the usual unitarity-defining requirement that inner products must be time independent, which is obviously only realized here when $\Ham$ is Hermitian. This signifies that the standard Dirac inner product employed above is the incorrect inner product for dealing with non-Hermitian Hamiltonians. Though it is possible to define a unitary inner product based directly on the anti-linear $\PT$ symmetry of a Hamiltonian\footnote{This inner product is defined by identifying an additional symmetry that is present in theories with unbroken $\PT$ symmetry that is related to the conservation of the $\PT$ ``charge'' ($\PT$ eigenvalue) of the Hamiltonian eigenstates \cite{Bender2004}. Though the this anti-linear-symmetry-based definition of the inner product is also perfectly valid, we embrace the definition based on pseudo-Hermiticity as it is more straightforwardly applied to the Heisenberg picture.}, we may also embrace the remarkable fact that every Hamiltonian that possesses an anti-linear symmetry and a bi-orthonormal system of eigenstates as in \eqref{biortho}, is necessarily \textit{pseudo-Hermitian} \cite{Mostafazadeh2002a,Mostafazadeh2002c}. Pseudo-Hermitian theory has been studied since the early days of QFT with the work of Dirac and Pauli \cite{Pauli1943} and was also considered by Lee, Wick, and their contemporaries as a general way to deal with indefinite inner product metrics \cite{Sudarshan1961,Lee1969a}. In this language, we can express the anti-linear symmetry of a Hamiltonian in terms of an invertible Hermitian operator $\IPo$, as
\begin{align} \label{HamPT}
\Ham = \IPo^{-1}\Ham^\dag\IPo \equiv \Ham^\psc \,,
\end{align}
where we have introduced the shorthand ``staurogram'' symbol $\psc$ to denote pseudo-Hermitian-conjugation. Here, $\IPo:\FS\to\FS$ is a linear Hermitian automorphism on the Hilbert space $\FS$ associated with the Hamiltonian $\Ham$. The pseudo-Hermiticity operator, $\IPo$, thus generates an anti-linear symmetry of the Hamiltonian in the sense of \eqref{HamPT}. For the present work, this anti-linear symmetry is that of the combined fundamental discrete $Z_2$ parity and time symmetries ($\PT$), but it is important to recall that this construction can also be applied to arbitrary anti-linear symmetries of a Hamiltonian\footnote{Additional technical details which are not directly relevant for this work regarding the pseudo-Hermitian formalism and how it relates to anti-linear symmetries may be found in \cite{Mostafazadeh2002a,Mostafazadeh2002b,Mostafazadeh2002c,Mostafazadeh2010}.}.

In light of the above, we may define our desired generalized inner product at the operator level, between two arbitrary states $\ket{a}$ and $\ket{b}$, as
\begin{align} \label{psip}
\psbraket{b}{a} \equiv \mel{b}{\IPo}{a} \,.
\end{align}
This new inner product also defines the left vacuum and left Hamiltonian eigenstates not as the Dirac-adjoints of their right side counterparts $\ket{0}$ and $\ket{\psi_n}$, but rather as the $\IPo$-adjoints $(\ket{0})^\psc=\bra{0}\!\IPo$ and $(\ket{\psi_n})^\psc=\bra{\psi_n}\!\IPo$. With this, the expression \eqref{badHE} is upgraded to
\begin{align} \label{goodHE}
\partial_t\!\mel{\psi_n}{\IPo}{\psi_m} = -i\mel{\psi_n}{\big(\IPo\Ham - \Ham^\dag\IPo\big)}{\psi_m} = 0 \,,
\end{align}
thus preserving time independence for a Hamiltonian that satisfies \eqref{HamPT}. This also implies that unitary time evolution, generated as usual by the operator $U(t)=e^{-i\Ham t}$, must be established through the relation
\begin{align} \label{tevo}
U^\psc U = \IPo^{-1} e^{i\Ham^\dag t} \IPo e^{-i\Ham t} = U^{-1}U = 1 \,.
\end{align}
The usual requirement that general observables $O$ be Hermitian is also upgraded to $O=O^\psc$ in this framework.

As previously mentioned, the strict reality of Hamiltonian eigenvalues in the anti-linear symmetry language is proven by demonstrating that all eigenstates of $\Ham$ are also eigenstates of $A$. This proof of reality in the pseudo-Hermitian language, as laid out in \cite{Mostafazadeh2002b}, relies on the existence of an invertible operator $\hIPo$ that transforms the Hamiltonian into a Hermitian form:
\begin{align}
\bar{\Ham} = \hIPo\Ham\hIPo^{-1} = \bar{\Ham}^\dag \,.
\end{align}
Since $\hIPo$ generates a similarity transformation, the eigenvalues of $\Ham$ and $\bar{\Ham}$ are necessarily equivalent while their eigenstates are related by $|\bar{\psi}_n\rangle=\hIPo\ket{\psi_n}$. The operator $\hIPo$ thus also maps the Hilbert space associated with $\Ham$ to a space where the appropriate inner product is simply the standard Dirac one. This further implies that inner products taken in each space are related by $\braket{\bar{b}}{\bar{a}}=\psbraket{b}{a}$, which leads to an important relationship between $\hIPo$ and the pseudo-Hermiticity of the original Hamiltonian:
\begin{align}
\IPo = \hIPo^\dag\hIPo \,.
\end{align}
It is in fact a general feature of pseudo-Hermitian Hamiltonians that their spectrum of eigenvalues will be real if and only if $\IPo$ admits such a decomposition \cite{Mostafazadeh2002b,Mostafazadeh2002c}.

It is also important to note that the familiar versions of the Wick contraction procedure and the Feynman contour prescription are also applicable in the pseudo-Hermitian framework. Wick's theorem does not depend on the inner product and the Feynman propagator is expressed as usual; for example, in a simple massive scalar theory we would have
\begin{align} \label{Feynprop}
D_F(x - y) \equiv \psmel{0}{\text{T}\,\phi(x)\phi(y)}{0} = \mel{0}{\IPo\,\text{T}\phi(x)\phi(y)}{0} = -i\int\frac{\dd^4p}{(2\pi)^4}\frac{e^{ip(x-y)}}{p^2 + m^2 - i\epsilon} \,,
\end{align}
where the only difference between this and the standard picture is that the matrix element that $D_F$ refers to is defined with respect to the inner product \eqref{psip}. Higher $n$-point correlators are also similarly defined with respect to the $\psIP$ inner product, however, this does not imply that the calculation of Feynman diagrams is unaffected (see e.g.\ \cite{Mannheim2018}). Indeed, as we will see shortly, the $\IPo$ operator enters directly into the definition of the S-matrix. 

Additionally, we remark on the fact that the present formalism also naturally allows for a spectral representation of two-point functions in an interacting fourth-order theory. Derivation of the spectral representation follows the same logic as in second-order theories, with the caveat that the relevant non-vanishing commutator is not between a field and its first derivative, but its third derivative \cite{Bender2008a}. This analysis leads to the conclusion that the usual spectral function cannot be positive definite in fourth-order theories, and though this would seem to lead to issues with unitarity, it has been shown that this positive-definiteness is not actually necessary in non-Hermitian theories \cite{Bender2002a}. Indeed, adoption of the $\PT$ inner product allows one to reconcile unitarity with spectral representation.

Finally we note that, generally speaking, while correlators may be calculated using path integral or interaction picture techniques, as usual, the former of these methods turns out to be the clearly preferable option. This is due to the simple fact that it is not always necessary to calculate $\IPo$ explicitly in the path integral formalism \cite{Jones2007}. However, as we will not perform any such calculations here, we will stick to the operator-based Heisenberg picture of QFT in what follows, as it forms the theoretical backbone that all other interpretations are based on.

\subsection{Unitarity in the Heisenberg picture} \label{subsec:uniheis}

We will employ the techniques of covariant quantization as laid out by Kugo, Nakanishi, and Ojima \cite{Nakanishi1990,Kugo1997}, which builds on the well-known formalism of Lehmann, Symanzik, and Zimmerman (LSZ) \cite{Lehmann1955,Glaser1957}. In this framework, one assumes that the fields in an action describing an interacting QFT, generally denoted here as $\phi(x)$, may be treated as free (non-interacting) fields in the (weak) limit of the asymptotic past or future:
\begin{align} \label{asymlim}
\phi(x) \to 
\left\{\begin{array}{ll}
	\phi^\text{in}(x) \,, & x^0 = t \to -\infty \\ 
	\phi^\text{out}(x) \,, & x^0 = t \to +\infty
\end{array}\right._{\;{\textstyle .}}
\end{align}
We will generally denote asymptotic fields as $\phi^\text{as}(x)$ when either the ``in'' or ``out'' limit may be applied. Taking all fields to be Hermitian, their asymptotic versions may be decomposed as a sum of products of c/a operators and plane wave functions:
\begin{align} \label{phioscdec}
\phi^\text{as}(x) = \sum_{\bm p}\Big(a^\text{as}(\bm{p})f_{\bm p}(x;m) + a^\text{as}_g({\bm p})g_{\bm p}(x;m) + \cdots + \hc\Big) \,.
\end{align}
Here, ${\bm p}$ represents the three-dimensional spatial components of the four-momentum $p^\alpha$, while $f_{\bm p}(x;m)$ and $g_{\bm p}(x;m)$ are single and double pole plane wave functions respectively, which satisfy increasing powers of the Klein-Gordon equation,
\begin{align}
\big(\Box - m^2\big)f_{\bm p}(x;m) = \big(\Box - m^2\big)^2g_{\bm p}(x;m) = 0 \,.
\end{align}
The ``$\cdots$'' in \eqref{phioscdec} represents the possibility that even more pairs of independent c/a operators and higher pole plane waves are present in a field's decomposition; the number of such pairs required is dictated by the equations of motion (EOMs) and depends on the particular theory/gauge choice at hand. With all of this, the actual quantization is established by imposing commutation relations between the fields that depend on invariant delta functions associated with the degree of the pole for that propagator, $D(x-y;m)$, $E(x-y;m)$, etc. For example, a field $\phi(x)$ with the simple-pole propagator $-i/(p^2+m^2)$ will respect the commutation relation
\begin{align} \label{phicom}
\Com{\phi(x)}{\phi(y)} = D(x-y;m) \qquad\text{where}\qquad D(x-y;m) = \sum_{\bm p}\big(f_{\bm p}(x;m)f_{\bm p}^*(y;m) - \hc\big) \,.
\end{align}
Naturally, the c/a operators are then also subject to commutation relations that satisfy the field commutators after each field is decomposed as in \eqref{phioscdec}. We have been intentionally brief in describing all of this machinery here to keep the discussion focused, as we will see it in action in the following section. We also refer the reader to \cite{Kuntz2022a} for deeper discussions and precise definitions.

Assuming the asymptotic limit and decomposing fields per the above is what allows us construct a well-defined theory of quantum states on the basis of a unique Poincaré-invariant vacuum $\ket{0}$, which satisfies $a^\text{as}(\bm{p})\ket{0}=0$ following the ``$+$'' definition in \eqref{pmvacs}. This choice in turn defines the single particle states
\begin{align}
\ket{a;\text{as}} \equiv a^{\text{as}\,\dag}(\bm{p})\ket{0} \,,
\end{align}
which occupy the Hilbert space of our theory, denoted as $\FS$. We will always assume that this space is ``asymptotically complete'', which is simply the statement that the present Hilbert space is equivalent to the Hilbert space of the asymptotic past and future,
\begin{align} \label{asympcomp}
\FS^\text{out} = \FS = \FS^\text{in} \,.
\end{align}
Only with this assumption are we able to consistently express operators as products of asymptotic fields.

It is important to note that the construction of a theory's Hilbert space and our assumption of asymptotic completeness does not rely on the definition of an inner product, meaning that everything discussed in this section so far applies to both Hermitian and pseudo-Hermitian QFT. The notion of an inner product is however crucial if we are to take expectation values of quantum operators and extract any physical meaning from a theory. Indeed, unitarity and the usual interpretation of quantum probability per the Born rule \eqref{Born} only follows from this whole construction if $\FS$ is equipped with a positive-definite inner product. For standard Hermitian theory this requirement looks like $\braket{a}{a}>0$, though it is important to recall that the dual Hilbert space i.e.\ the left vectors, are only defined with respect to the right vectors and after one defines an inner product. As we have already seen, we must employ our pseudo-Hermitian inner product in $\PT$-symmetric theory where
\begin{align} \label{PTposdef}
\psbraket{a;\text{as}}{a;\text{as}} > 0 \qquad\text{for any}\qquad \ket{a;\text{as}} \neq 0 \,,
\end{align}
meaning that our dual Hilbert space is populated by the $\IPo$-adjoints
\begin{align}
\psbra{a;\text{as}}\equiv\bra{0}a^\text{as}(\bm p)\IPo \,.
\end{align}

In the present language, physical interactions are understood in terms of free particles that exist in the asymptotic past, come together to scatter off one another, and continue on into the asymptotic future, once again as free particles. This transformation between in and out states through scattering events is described by the S-matrix operator $\Smat$:
\begin{align}
\phi^\text{out}(x) = \Smat^{-1}\phi^\text{in}(x)\Smat \qquad\text{or}\qquad \ket{a;\text{out}} = \Smat^{-1}\ket{a;\text{in}} \,, \qquad\text{where}\qquad \Smat\ket{0} = \ket{0} \,.
\end{align}
The square of matrix elements constructed from $\Smat$, $\Smat_{ab}=\!\tensor[_\eta]{\mel{b;\text{out}}{S}{a;\text{in}}}{}$, are what actually determine the transition probability of $\ket{a}\to\ket{b}$. As in the usual Hermitian theory, consistent interpretation of these transition probabilities is only possible if we make the assumption that $\Smat$ is pseudo-unitary which, per the S-matrix's definition in terms of the simultaneous $\pm\infty$ limit of the evolution operator in \eqref{tevo}, must take the form
\begin{align} \label{pseudouni}
\Smat^\psc \Smat = \Smat^{-1}\Smat = 1
\end{align}
as opposed to the traditional definition $\Smat^\dag\Smat=1$. It is important to reinforce that this assumption of pseudo-unitarity of the S-matrix may be made regardless of whether the related Hilbert space contains negative norm states; ghosts do not threaten this notion of unitarity, only the way in which one defines the interpretation of probability.

We may also consider a definition of the S-matrix given in terms of the completion of asymptotic in and out states. With the assumption of asymptotic completeness and the fact that it allows us to express any operator as a product of asymptotic states, this is straightforwardly given by
\begin{align}
\Smat = \sum_i\ket{a_i;\text{in}}\bra{a_i;\text{out}}\IPo \,,
\end{align}
which allows one to relate asymptotic eigenstates as
\begin{align}
\ket{a;\text{out}} = \Smat^\psc\ket{a;\text{in}} = \IPo^{-1}\Smat^\dag\IPo\ket{a;\text{in}} \,,
\end{align}
thus allowing for \eqref{asympcomp} to be expressed equivalently as the requirement that $\FS$ is invariant under action of the S-matrix,
\begin{align}
\Smat\FS^\text{as} = \FS^\text{as} \,.
\end{align}
Putting all of the above together, we may establish the relation
\begin{align} \label{unireq}
1 &= \psbraket{a;\text{as}}{a;\text{as}} = \mel{a;\text{as}}{\IPo\Smat^\psc\Smat}{a;\text{as}} = \mel{a;\text{as}}{\Smat^\dag\IPo\Smat}{a;\text{as}} \nnr
&= \sum_n\big(\mel{a;\text{as}}{\Smat^\dag\IPo}{n;\text{as}}\mel{n;\text{as}}{\IPo\Smat}{a;\text{as}}\big) = \sum_n\big|\psmel{n;\text{as}}{\Smat}{a;\text{as}}\big|^2 \,,
\end{align}
which defines a probabilistic notion of unitarity, distinct from the pseudo-unitarity requirement \eqref{unireq}, as the requirement that the probabilities of all possible outcomes for a given scattering event must sum up to unity.

Naturally, we may also establish a familiar expression of the optical theorem in the pseudo-Hermitian framework. Parameterizing the S-matrix as $\Smat=1+i\Tmat$, where all non-trivial interactions are described by the transfer matrix $\Tmat$, \eqref{pseudouni} implies
\begin{align}
\Tmat^\psc\Tmat &= i\big(\Tmat^\psc - \Tmat\big) \\
\sum_n\mel{g}{\Tmat^\dag\IPo}{n}\mel{n}{\IPo\,\Tmat}{f} &= i\left(\mel{g}{\Tmat^\dag\IPo}{f} - \mel{g}{\IPo\,\Tmat}{f}\right) \,.
\end{align}
Here, we have formed a matrix element of the first line using the $\psIP$ inner product between two arbitrary states $\ket{f}$ and $\ket{g}$, and inserted the unit operator $1=\sum_n\ket{n}\bra{n}\IPo$ to arrive at the second line. This version of the optical theorem carries the same diagrammatic interpretation and implies the same cutting rules as it does in the Hermitian picture, irrespective of the appearance of $\IPo$ in the matrix elements; the sum over all possible cuts that split a loop diagram into tree-level diagrams is proportional to the imaginary part of that loop diagram amplitude. We recall that Feynman integrals are set up and evaluated in the same way as in Hermitian theory, meaning that perturbative unitarity may also be established by calculating diagrams without passing to the asymptotic limit, in the standard way (see e.g. \cite{Mannheim2018}).

\subsection{Physical subspace in gauge theories} \label{subsec:physsub}

Up until this point we have assumed that we are working with a positive-definite Hilbert space satisfying \eqref{PTposdef}, however, we must also consider the realistic scenario that there exists states in $\FS$ that violate this condition (states with zero or negative norm). In this case we say that $\FS$ is an indefinite-metric Hilbert space. Such spaces unavoidably arise when working with gauge theories and it is crucial that we are able to identify the transverse physical subspace $\bFSp\subset\FS$ where
\begin{align}  \label{pdbFSp}
\psbraket{a}{a} > 0 \qquad\text{for all}\qquad \ket{a} \in \bFSp \,,
\end{align}
if we are to take any meaning from scattering events in the theory. It is only with a clear definition of $\bFSp$ that we may establish unitarity through the relation \eqref{unireq} with respect to the transverse physical S-matrix between states in $\bFSp$.

As established by Nakanishi and collaborators \cite{Nakanishi1990}, consistent definition of this positive-definite subspace rests on the assumption of asymptotic completeness in the total Hilbert space \eqref{asympcomp}, the pseudo-unitarity of the complete S-matrix on the total Hilbert space \eqref{pseudouni} with respect to the inner product \eqref{psip}, and on the existence of a physical subspace $\FSp$ that is asymptotically complete ($\FSp^\text{out}=\FSp^\text{in}$) and positive semi-definite with respect to the (indefinite) inner product on the total space:
\begin{align}  \label{psdFSp}
\psbraket{a}{a} \geq 0 \qquad\text{for all}\qquad \ket{a} \in \FSp \,.
\end{align}
We follow the convention of Nakanishi and refer to $\FSp$ as the ``physical'' Hilbert space, though it should not be confused with the ``transverse physical'' Hilbert space of genuine physical particles, $\bFSp=\FSp/\FS_0$ in \eqref{pdbFSp}, where $\FS_0$ is the zero norm subspace of $\FSp$. 

In order to define the positive semi-definite subspace $\FSp$, we must enforce a ``subsidiary condition'' on $\FS$ to identify the states that satisfy \eqref{psdFSp}. Though selecting a satisfactory subsidiary condition is in general a very complex task that has received much attention from theorists in the past, a general and comprehensive method has been laid out by Becchi-Rouet-Stora-Tyutin (BRST) that revolves around the introduction of so-called BRST symmetry \cite{Becchi1975,Becchi1976,Fradkin1970}. This work has proved indispensable for establishing rigorous proofs of stability, renormalizability, and unitarity in the Standard Model, and we will employ it for our purposes as well.

To establish a BRST symmetry in a given gauge theory, one must select gauge-fixing conditions $G^a$ for each gauge symmetry of the action and introduce a set of unphysical fields that serve to enforce the gauge conditions and precisely cancel unphysical contributions to loop diagrams at every order that arise from the non-gauge-invariant parts of the original fields. These new fields are the so-called Nakanishi-Lautrup (NL) fields $b_a$ and the Faddeev-Popov (FP) ghosts and anti-ghosts $c^a$ and $\bar{c}_a$ (not to be confused with the Ostrogradsky ghosts of the ghost problem). Crucially, despite the fact that we will encounter a non-Hermitian action/Hamiltonian, we will assume here that \textit{all fields are Hermitian}\footnote{This does not preclude the possibility of applying this formalism to a complex Higgs doublet for example, rather, we assume that all fields have the same behavior under Hermitian conjugation that they would in standard theory.}, just as they are in the standard theory. It is also important to note that BRST symmetry is global and graded in terms of each type of fields ``ghost number''; the original bosons and NL fields carry ghost number 0 while the ghosts and anti-ghosts carry ghost number 1 and $-1$ respectively.

BRST transformations are generated by the nilpotent operator $\BChg$ which acts on each type of field according to
\begin{align}
\begin{aligned} \label{BRSTtrans}
&\delta\phi_a = \sum_{\xi}\Big(\delta^{(\xi)}\phi_a\Big)\Big|_{\xi^b=c^b} \eqspace& &\delta b_a = 0 \\
&\delta c^a = c^b\partial_bc^a& &\delta\bar{c}_a = ib_a \,,
\end{aligned}
\end{align}
where $\delta X=i\Com{\BChg}{X}_\mp$ and $\delta^{(\xi)}\phi_a$ represents the gauge transformation of $\phi_a$ with respect to the local gauge transformation parameter $\xi^a(x)$. With this, we have enough information to construct the total BRST action
\begin{align} \label{genST}
\Act_\text{T} = \Act + \Act_\text{BRST} \,,
\end{align}
where $\Act$ is the original gauge-invariant action of the theory and the BRST fields enter through
\begin{align}
\Act_\text{BRST} = -i\int\dx\delta\big(\bar{c}_aG^a\big) \,,
\end{align}
which serves to gauge-fix the original theory and provide terms for the NL fields and FP ghosts. The total action \eqref{genST} is guaranteed to be BRST-invariant, with $\Act$ and $\Act_\text{BRST}$ each being invariant on their own, as a result of \eqref{BRSTtrans} and the nilpotency of $\BChg$. One of the most important aspects of the BRST setup is thus made manifest; the total action is able to maintain information about the original gauge symmetry through the action of $\BChg$ on the $\phi_a$, even after being gauge-fixed. This feature is in the end what allows for the identification of physical gauge-invariant subspace of the total Hilbert space.

The precise way in which the BRST construction leads to the cancellation of unphysical DOFs is known as the Kugo-Ojima (KO) quartet mechanism \cite{Kugo1978,Kugo1978a,Kugo1979a,Kugo1979}. After quantization following the LSZ procedure laid out in the previous section, we are left with a Hilbert space $\FS$ that contains both physical and unphysical states, the former of which are identified by the subsidiary condition
\begin{align} \label{subsid}
\BChg\ket{a} = 0 \qquad\text{for all}\qquad \ket{a} \in \FSp \,.
\end{align}
The subspace $\FSp$ is thus identified as the space populated by all $\ket{a}$ that are BRST-invariant, which is necessarily positive semi-definite owing to the lack of all non-gauge-invariant states. With this, we may make the final identification of the desired positive-definite subspace as
\begin{align}
\bFSp = \text{Ker}\BChg/\text{Im}\BChg
\end{align}
known as the BRST cohomology space (the transverse physical subspace) where $\text{Ker}\BChg=\FSp$ and $\text{Im}\BChg=\BChg\FS=\FS_0$. With the removal of all zero norm states that populate $\FS_0$, the resulting $\bFSp$ is then found to be positive-definite with all of its states satisfying \eqref{PTposdef}, thus allowing for unitarity of the transverse physical S-matrix on $\bFSp$ to be established per \eqref{unireq}. 

A well-written proof of the Hermitian quartet mechanism may be found in \cite{Nakanishi1990} and we will not go over each step here, however, it is important that we sketch out its most important details to see how the proof may be easily adapted to the pseudo-Hermitian picture. It is possible to quantify precisely which states present in the total $\FS$ occupy the subspaces $\FSp$, $\FS_0$, and thus also $\bFSp$, based solely on representations of the BRST algebra. Due to the nilpotency of $\BChg$, all states in $\FS$ are either annihilated by $\BChg$ (satisfying the subsidiary condition \eqref{subsid}), or are transformed into another state in $\FS$ that is in turn annihilated by $\BChg$. BRST-invariant states that are not related to any other states in $\FS$ through BRST transformation are called singlets, while states that are not BRST-invariant (called parents) or may be generated through the BRST transformation of another state (called daughters), necessarily come in doublet pairs that we denote as $\ket{\pi}$ and $\ket{\delta}$, respectively. Additionally, since the daughter states necessarily have zero norm, non-degeneracy of $\FS$ and the requirement that the total action must carry ghost number zero implies that for each doublet there must also exist another doublet with conjugate ghost numbers. These pairs of doublets constitute the titular quartets whose familial relationships are characterized by the relations
\begin{align} \label{pdrels}
&\BChg\big|\pi^{(0)}\big\rangle = \big|\delta^{(1)}\big\rangle \neq 0& &\BChg\big|\pi^{(-1)}\big\rangle = \big|\delta^{(0)}\big\rangle \neq 0 \,,
\end{align}
where superscripts indicate FP ghost number.

The central point of the quartet mechanism is to guarantee that only genuine physical particles are able to appear as external asymptotic states in scattering events. This occurs as a result of the facts: i) physical states are always orthogonal to all unphysical states and ii) all states classified as a member of a quartet necessarily appear only in zero norm combinations i.e.\ they are confined \cite{Nakanishi1990}. The proof of this confinement follows after a demonstration that all states in $\FS$ can be categorized as either BRST singlets or members of a quartet that are related by \eqref{pdrels} and subject to the quartet-defining relation
\begin{align} \label{quartetdef}
\tensor[_\IPo]{\big\langle\delta^{(0)}\big|\pi^{(0)}\big\rangle}{} = \tensor[_\IPo]{\big\langle\pi^{(-1)}\big|\BChg\big|\pi^{(0)}\big\rangle}{} = \tensor[_\IPo]{\big\langle\pi^{(-1)}\big|\delta^{(1)}\big\rangle}{} \neq 0 \,.
\end{align}
This is the natural extension of the Hermitian theory definition of a BRST quartet where the usual inner product is replaced by the pseudo-Hermitian inner product \eqref{psip}. We claim that the most straightforward way to ensure that this relation holds for general gauge theories in the pseudo-Hermitian framework is to define $\IPo$ on the basis of the transverse physical Hamiltonian alone. With this, $\IPo$ may be constructed solely from products of states in $\bFSp$ and thus commutes with all quartet c/a operators as well as $\BChg$. The quartet mechanism and proof of confinement then follows just as it does in the Hermitian theory since all inner products between quartet members are equivalent whether one employs the Dirac inner product $\langle\cdot\rangle$ or the pseudo-Hermitian inner product $\psIP$.

In summary, the overall path to a demonstration of unitarity goes as follows. After establishing a BRST symmetry, identifying the physical subspace with the BRST subsidiary condition \eqref{subsid}, and removing all zero norm states belonging to quartets from that subspace, we are left with a subspace $\bFSp$ of genuine physical states where unitarity rests on the positivity of the inner product. In Hermitian QFT, this positivity is guaranteed for a theory with healthy particles only, but is violated when Ostrogradsky ghosts are present. However, as we have seen in our toy model, this aspect of the ghost problem may be avoided via complex deformation to a $\PT$-symmetric non-Hermitian theory and adoption of the pseudo-Hermitian inner product \eqref{PTposdef}. The assertion that $\IPo$ commutes with $\BChg$ and all of the quartet c/a operators allows the unitarity-defining mechanisms in the gauge and ghost regimes to work in harmony and establish an interpretation of probability in the theory as a whole.

\section{Quadratic Gravity} \label{sec:QG}

\subsection{Gauge-fixed total action}

With all of the prerequisites established, we can begin our application of the $\PT$-symmetric/ pseudo-Hermitian framework to quadratic gravity. The classical theory is described by the most general action containing all the independent squares of the Riemann tensor, which may be parameterized as
\begin{align}
\Act &= \int\dx\sqrt{-g}\bigg[\Mp^2R - \frac{1}{2\alpha_g^2}C_{\alpha\beta\gamma\delta}C^{\alpha\beta\gamma\delta} + \beta R^2\bigg] \,,
\end{align}
where $\alpha_g$ and $\beta$ are arbitrary dimensionless coupling constants, $R$ the Ricci scalar, and $C$ is the Weyl tensor. To get this action into a more workable form, we may subtract a total derivative proportional to the Gauss Bonnet invariant, $\mathcal{G} = C_{\alpha\beta\gamma\delta}C^{\alpha\beta\gamma\delta} - 2R_{\alpha\beta}R^{\alpha\beta} + \frac{2}{3}R^2$, as it does not contribute to the EOMs or propagators. Additionally, as we will be focused on the ghost problem that arises in the spin-2 sector of the theory, we will set $\beta=0$ simply to avoid clutter, since the corresponding $R^2$ contributes only additional non-ghostly scalar DOFs to the theory. After also defining $m^2=\alpha_g^2\Mp^2$, we are left with the action that will serve as the starting point for our analyses,
\begin{align} \label{AQG}
\Act_\text{QG} = \int\dx\sqrt{-g}\bigg[\frac{1}{\alpha_g^2}\bigg(m^2R - R_{\alpha\beta}R^{\alpha\beta} + \frac{1}{3}R^2\bigg)\bigg] \,.
\end{align}
We also note that this version of the general quadratic gravity action without $\beta R^2$ corresponds to a low-energy realization of Weyl's conformal gravity. This theory generally contains only the square of the Weyl tensor in the gravitational action, but generally also includes a non-minimally coupled matter scalar (dilaton) whose vacuum expectation value can be identified with the Planck mass, thus generating an Einstein-Hilbert term after the spontaneous breakdown of conformal symmetry \cite{Mannheim2012}.

The action \eqref{AQG} is invariant under the diffeomorphisms
\begin{align} \label{metdiff}
g'_{\alpha\beta} = \met + \frac{\alpha_g}{m}\Lag_\xi\met = \met + \frac{\alpha_g}{m}\big(\partial_\gamma\met + g_{\alpha\gamma}\partial_\beta + g_{\beta\gamma}\partial_\alpha\big)\xi^\gamma
\end{align}
where $\Lag_\xi$ is the Lie derivative in the direction of an arbitrary vector $\xi^\alpha(x)$. To fix the associated gauge freedom, we select the de Donder-style condition
\begin{align} \label{gaugecon}
G^\alpha = \frac{m}{\alpha_g}g^{\alpha\beta}g^{\gamma\delta}\bigg(\partial_\delta g_{\beta\gamma} - \frac12\partial_\beta g_{\gamma\delta}\bigg) + \frac12g^{\alpha\beta}b_\alpha \,.
\end{align}
As discussed in Section \ref{subsec:physsub}, we will implement the condition $G^\alpha=0$ by establishing a BRST symmetry through the introduction of an NL field $b_\alpha(x)$, FP ghost $c^\alpha(x)$, and FP anti-ghost $\bar{c}_\alpha(x)$ that transform under BRST according to
\begin{align}
\delta b_\alpha = 0 \eqspace \delta c^\alpha = \frac{\alpha_g}{m}c^\beta\partial_\beta c^\alpha \eqspace \delta \bar{c}_\alpha = ib_\alpha \,.
\end{align}
Following \eqref{BRSTtrans}, the metric must also transform under BRST as the Lie derivative in the direction of $c^\alpha$,
\begin{align} \label{nonlintrans}
\delta\met = \frac{\alpha_g}{m}\big(\partial_\gamma\met + g_{\alpha\gamma}\partial_\beta + g_{\beta\gamma}\partial_\alpha\big)c^\gamma \,.
\end{align}
With this, the total BRST action is given by
\begin{align} \label{AT}
\Act_\text{T} = \Act_\text{QG} + \Act_\text{BRST}
\end{align}
where the condition \eqref{gaugecon} is enforced by the NL field $b^\alpha$ that, along with the FP ghost and anti-ghost, enters into the total action through
\begin{align} \label{ABRST}
\Act_\text{BRST} &= - i\int\dx\gdet\,\delta\Big(\bar{c}_\alpha\,G^\alpha\Big) \nnr
&= \int\dx\gdet\bigg[b_\alpha\bigg(\frac{m}{\alpha_g}\Gamma^{\alpha\beta}{}_\beta+\frac12b^\alpha - \frac{i\alpha_g}{m}\bar{c}_\beta\nabla^{(\alpha}c^{\beta)}\bigg) \nnr
&\phantom{= \int\dx\gdet\bigg[\,}+ i\bar{c}_\alpha\Big(\nabla_\beta\nabla^\beta c^\alpha + c_\beta R^{\alpha\beta} - 2\nabla_\gamma c_\beta\Gamma^{\alpha\beta\gamma}\Big)\bigg]\,.
\end{align}

The dynamical part of the gravitational field that will actually determine the asymptotic quantum states in our theory corresponds to tensor perturbations $\yab(x)$ that we expand around the classical background metric. Taking the background to be Minkowski ($\met=\mmet$), we may thus derive an action for $\yab$ by applying the expansion
\begin{align} \label{ylin}
\met \trans \mmet + \frac{\alpha_g}{m}\yab
\end{align}
to \eqref{AT}. This expansion allows us to separate the gravitational part of the action as
\begin{align}
\Act_\text{QG} = \Act_\text{QG}^{(0)} + \Act_\text{QG}^{(\text{int})} \,,
\end{align}
where $\Act_\text{QG}^{(0)}$ is the free (quadratic in $\yab$) part and $\Act_\text{QG}^{(\text{int})}$ contains an infinite tower of momentum-dependent $\Ord[\alpha_g]$ interaction terms. The free part will be most important for our current purposes and may be written as
\begin{align} \label{AQG0}
\Act_\text{QG}^{(0)} = \frac14\int\dx\iyab\bigg(\mathcal{E}_{\alpha\beta\gamma\delta} - \frac{1}{m^2}\mathcal{W}_{\alpha\beta\gamma\delta}\bigg)\psi^{\gamma\delta}
\end{align}
after integrating by parts and neglecting the resulting dynamically irrelevant boundary terms. Here, the kinetic operators $\mathcal{E}$ and $\mathcal{W}$ correspond to the Einstein-Hilbert and Weyl parts of the action \eqref{AQG}, respectively, and are defined as
\begin{align}
&\mathcal{E}_{\alpha\beta\gamma\delta} \equiv \frac12\big(\eta_{\alpha\gamma}\eta_{\beta\delta} + \eta_{\alpha\delta}\eta_{\beta\gamma}\big)\Box - \eta_{\alpha\beta}\eta_{\gamma\delta}\Box + \eta_{\alpha\beta}\partial_\gamma\partial_\delta + \eta_{\gamma\delta}\partial_\alpha\partial_\beta \nnr
&\phantom{\mathcal{E}_{\alpha\beta\gamma\delta} \equiv} - \frac12\big(\eta_{\alpha\gamma}\partial_\beta\partial_\delta + \eta_{\alpha\delta}\partial_\beta\partial_\gamma + \eta_{\beta\gamma}\partial_\alpha\partial_\delta + \eta_{\beta\delta}\partial_\alpha\partial_\gamma\big) \\
&\mathcal{W}_{\alpha\beta\gamma\delta} \equiv \frac12\big(\eta_{\alpha\gamma}\eta_{\beta\delta} + \eta_{\alpha\delta}\eta_{\beta\gamma}\big)\Box^2 - \frac13\eta_{\alpha\beta}\eta_{\gamma\delta}\Box^2 + \frac13\big(\eta_{\alpha\beta}\partial_\gamma\partial_\delta + \eta_{\gamma\delta}\partial_\alpha\partial_\beta\big)\Box \nnr
&\phantom{\mathcal{W}_{\alpha\beta\gamma\delta} \equiv} - \frac12\big(\eta_{\alpha\gamma}\partial_\beta\partial_\delta + \eta_{\alpha\delta}\partial_\beta\partial_\gamma + \eta_{\beta\gamma}\partial_\alpha\partial_\delta + \eta_{\beta\delta}\partial_\alpha\partial_\gamma\big)\Box + \frac23\partial_\alpha\partial_\beta\partial_\gamma\partial_\delta \,, \label{WeylOp}
\end{align}
where $\Box=\partial_\alpha\partial^\alpha$. The gauge invariance of the full action manifests through the linearization of \eqref{metdiff} at zeroth-order in $\alpha_g$:
\begin{align} \label{ytrans}
\yab' = \yab + \partial_\alpha\xi_\beta + \partial_\beta\xi_\alpha \,.
\end{align}

The free part of the BRST action \eqref{ABRST} is also obtained by performing the linearization \eqref{ylin} and dropping the $\Ord[\alpha_g]$ interaction terms:
\begin{align} \label{ABRST0}
\Act_\text{BRST}^{(0)} = \int\dx\bigg[b_\alpha\bigg(\partial_\beta\iyab - \frac12\partial^\alpha\psi_\beta{}^\beta + \frac12b^\alpha\bigg) + i\bar{c}_\alpha\Box\,c^\alpha\bigg] \,.
\end{align}
With this, the entire free total action is simply
\begin{align} \label{AT0y}
\Act_\text{T}^{(0)} = \Act_\text{QG}^{(0)} + \Act_\text{BRST}^{(0)} \,,
\end{align}
where the invariance of $\Act_\text{QG}^{(0)}$ under the transformation \eqref{ytrans} is violated by the presence of $\Act_\text{BRST}^{(0)}$. The gauge symmetry is however still encoded in the theory by its invariance under the linearized BRST transformations
\begin{align} \label{ylintrans}
&\delta\yab = \partial_\alpha c_\beta + \partial_\beta c_\alpha& &\delta b_\alpha = 0& &\delta c^\alpha = 0& &\delta\bar{c}_\alpha = ib_\alpha \,,
\end{align}
so that $\delta\Act_\text{T}^{(0)}=0$. 

It is also important to note that the $\psi$--$\psi$ propagator that results from the action \eqref{AT0y} takes the general form
\begin{align} \label{yprop}
\psmel{0}{\psi\psi}{0} \sim -\frac{im^2}{p^2(p^2 + m^2)} = -\frac{i}{p^2} + \frac{i}{p^2 + m^2} \,.
\end{align}
This double pole structure indicates the presence of both massless and massive DOFs contained within the field $\yab$. In standard Hermitian theory, the relative minus between the poles signifies that the massive part must correspond to a ghost, however, as we shall see shortly, this is only the case when $\yab$ is Hermitian.

\subsection{Complex deformation in quadratic gravity} \label{subsec:compdefQG}

Our next task is to reinterpret the fourth-order theory described above as a second-order non-Hermitian theory, just as we did for the Pais-Uhlenbeck toy model in Section \ref{subsec:compdef}. Anticipating the effect of introducing an auxiliary field and rotating it into the complex plane as in \eqref{uytrans}, we establish the complex off-diagonal second-order action 
\begin{align} \label{Aaux0}
\Act_\text{aux}^{(0)} = \frac14\int\dx\bigg[\Big(\iyab - 2i\iHab\Big)\mathcal{E}_{\alpha\beta\gamma\delta}\psi^{\gamma\delta} - m^2\Big(\Hab\iHab - \Hs{\alpha}\Hs{\beta}\Big)\bigg] \,,
\end{align}
where $\Hab(x)$ is a real auxiliary field analogous to the toy model $y(t)$. As required, this action is equivalent to the fourth-order action \eqref{AQG0} after integrating out the auxiliary field with its EOM
\begin{align} \label{HEOM}
\Hab = -\frac{i}{m^2}\bigg(\mathcal{E}_{\alpha\beta\gamma\delta} - \frac13\mmet\mathcal{E}_{\mu}{}^{\mu}{}_{\gamma\delta}\bigg)\psi^{\gamma\delta} \,.
\end{align}
This new auxiliary action also maintains the same gauge symmetry under linearized diffeomorphisms as its fourth-order counterpart, with $\yab$ transforming as in \eqref{ytrans} and $\Hab$ being invariant:
\begin{align}
\Hab' = \Hab \,.
\end{align}

The factor of $i$ appearing in \eqref{HEOM} clearly implies that $\yab$ is a non-Hermitian field and we may thus decompose it as
\begin{align} \label{ydef}
\yab = \hab + i\Hab \,,
\end{align}
where we have defined $\hab(x)$ to be the real part and taken $\Hab(x)$ to correspond to the imaginary part. Under this decomposition, the free gravitational action \eqref{AQG0} becomes
\begin{align} \label{Adiag0}
\tilde{\Act}_\text{QG}^{(0)} = \frac14\int\dx\bigg[h^{\alpha\beta}\mathcal{E}_{\alpha\beta\gamma\delta}h^{\gamma\delta} + H^{\alpha\beta}\mathcal{E}_{\alpha\beta\gamma\delta}H^{\gamma\delta} - m^2\Big(\Hab\iHab - \Hs{\alpha}\Hs{\beta}\Big)\bigg] \,,
\end{align}
where we note that the off diagonal $h$--$H$ kinetic terms cancel exactly, leaving us with an action that is simply a sum of the linearized Einstein-Hilbert action and the massive spin-2 Fierz-Pauli action, with no relative minus sign that would traditionally signal the presence of a ghost. Moreover, since both $\hab$ and $\Hab$ are real fields, this free action is in fact Hermitian. This is not true of the infinite tower of interaction terms that constitute $\tilde{\Act}_\text{QG}^{(\text{int})}$ however, where every term containing $(H)^n$ with odd $n$ appears with a factor of $i$.

An additional consideration is in order with regard to the BRST terms in the total action after performing our complex deformation of the original theory. Applying the expansion \eqref{ydef} directly to \eqref{ABRST0} will yield a complex linearized BRST action that couples both $\hab$ and $\Hab$ to the BRST fields. However, the preferable situation is to have only $\hab$ couplings present (at least at the free level) since it is the only field that varies under a gauge (BRST) transformation and having additional $\Hab$ couplings will only complicate the structure of propagators down the road. The simple way around this scenario is to ``Hermitize'' the BRST contributions by replacing \eqref{ABRST} as
\begin{align} \label{Hermitize}
\Act_\text{BRST} \trans \tilde{\Act}_\text{BRST} = \frac12\Big(\Act_\text{BRST} + \big(\Act_\text{BRST}\big)^\dag\Big) \,.
\end{align}
This has the effect of removing $\Hab$ from the free part of the BRST action, which is now given by
\begin{align}
\tilde{\Act}_\text{BRST}^{(0)} = \int\dx\bigg[b_\alpha\bigg(\partial_\beta h^{\alpha\beta} - \frac12\partial^\alpha h_\beta{}^\beta + \frac12b^\alpha\bigg) + i\bar{c}_\alpha\Box\,c^\alpha\bigg] \,,
\end{align}
though we note that even powers of $\Hab$ still appear in the interactions terms contained in $\tilde{\Act}_\text{BRST}^{(\text{int})}$. With this, the free BRST total action is given by 
\begin{align} \label{AT0}
\tilde{\Act}_\text{T}^{(0)} = \tilde{\Act}_\text{QG}^{(0)} + \tilde{\Act}_\text{BRST}^{(0)} \,,
\end{align}
which is invariant under the BRST transformations \eqref{ylintrans} and
\begin{align}
\delta h_{\alpha\beta} = \partial_\alpha c_\beta + \partial_\beta c_\alpha \eqspace \delta H_{\alpha\beta} = 0 \,.
\end{align}

Thus, after the complex deformation of linearized quadratic gravity we are left with second-order diagonal theory with a Hermitian free action and non-Hermitian interaction terms. A similar kind of separation of the spin-2 DOFs in quadratic gravity has been performed for the Hermitian form of the theory in \cite{Kuntz2022b}, which naturally looks nearly the same as it does here, except that it contains no factors of $i$ in the interaction terms and carries an overall minus sign in the $\Hab$ part of the free action that indicates $\Hab$'s role as a ghost. Here we have instead simply introduced $i\Hab$ instead of $\Hab$, which has the effect of canceling the overall minus and making $\Hab$ behave as a healthy spin-2 field with complex interactions. As we have already discussed in some detail, this may still constitute a sensible physical theory if we are able to realize an anti-linear symmetry under $\PT$ transformation. Such a symmetry is easily established with the transformation properties
\begin{align} \label{fieldPT}
\begin{array}{c|cccccc}
& \met & \hab & \Hab & b_\alpha & c^\alpha & \bar{c}_\alpha \\
\hline
\Par & - & - & - & - & -  & - \\
\Tim & - & - & + & - & -  & -
\end{array}_{\;{\textstyle ,}}
\end{align}
where each field transforms in the standard way as $\Par$ and $\Tim$ pseudo-scalars, with the exception of $\Hab$ which we take to be a scalar under time inversion. With these properties, $i\Hab$ is a $\PT$ scalar (as are all the other fields without factors if $i$) which guarantees that all $\Ord[\alpha_g]$ interactions terms will be $\PT$-invariant.

It is also important to address the connections between the present theory and accepted gravitational physics. As mentioned in the Introduction, it is already well-established that all of the classical observations predicted by GR can also be reproduced from the classical QG action \eqref{AQG} up to the necessary precision. This is thanks to the presence of QG's Einstein-Hilbert term, despite the differences between Hermitian GR and pseudo-Hermitian QG that appear at the quantum level related to wavefunction normalization and the modified inner product. This follows from our introduction of the complex deformation at the level of perturbations around the classical background metric. With this, it is straightforward to realize all of the important low-energy classical observables predicted by GR, provided that $\Hab$ is heavy enough to evade experimental bounds on massive gravity. This is indeed the most realistic case when $\Hab$ acquires its mass as the result of the spontaneous breaking of scale symmetry for example, where $m\approx10^{-2}\Mp$ \cite{Kuntz2022c}.

There also exists an important relationship between between the standard and non-Hermitian theories at the quantum level, where the advantages of considering the non-Hermitian picture are actually manifest. The non-zero propagators in the second-order theory take the general forms
\begin{align}
\psmel{0}{hh}{0} \sim -\frac{i}{p^2} \eqspace \psmel{0}{HH}{0} \sim - \frac{i}{p^2 + m^2} \,,
\end{align}
which is consistent with a decomposition of the $\psi$--$\psi$ propagator \eqref{yprop} per \eqref{ydef}:
\begin{align} \label{yhHprop}
\psmel{0}{\psi\psi}{0} \sim \psmel{0}{hh}{0} - \psmel{0}{HH}{0} \,.
\end{align}
Indeed, the fourth-order theory presented in the last section maintains the important Pauli-Villars-style propagator that allowed Stelle to establish the renormalizability of QG in \cite{Stelle1977}, despite the fact that that study employed the Dirac inner product and a Hermitian $\yab$. On the other hand, the present second-order formulation lends itself to the analysis of unitarity in the theory, as it allows us to see that the relative minus signs in \eqref{yprop} and \eqref{yhHprop} need not be related to a negative $\Hab$ norm if we allow for $\yab$ to be complex.

To conclude this section, we should also comment on the relationship between the present model and the related work of Lee and Wick on higher derivative theories \cite{Lee1969a,Lee1970}. These models also contain a physical DOF that acts as a UV regulator, however, since they are based on standard Hermitian theory, the regulating field is a ghost with negative norm and unitarity is claimed to follow from a modified Feynman contour prescription\footnote{This modified $i\epsilon$-prescription has received criticism due to its inherent violation of Lorentz invariance \cite{Nakanishi1971,Lee1971,Nakanishi1971a} and the fact that it is tricky to define non-perturbatively \cite{Boulware1984}.}. In this setup, one also finds that radiative corrections generate a complex mass for the ghost. Despite the fact that the Feynman prescription in the present pseudo-Hermitian framework is unaltered, one may still expect that complex masses will also arise in our theory due the complex interaction terms. Though loop-order corrections do not lead to complex masses in fourth-order pseudo-Hermitian theory with $\phi^4$ interactions (see \cite{Mannheim2018}), it remains to be seen if this behavior is also realized in the present theory with its more complicated structure of interactions. Such calculations are beyond the scope of this work, but are certainly warranted in the future.

\subsection{Covariant quantization}

Moving on, we proceed with our investigation of unitarity by quantizing our second-order theory. This begins with a derivation of the matrix of propagators between each member of the full set of fields, $\Phi_A=\{\hab,\Hab,b_\alpha,c_\alpha,\bar{c}_\alpha\}$, by taking the inverse of the Hessian of the free action \eqref{AT0} in momentum space:
\begin{align} \label{propmat}
-i\mathcal{D}(p)^{AB} &= \left(\int\dx\frac{\delta^2\tilde{\Act}_\text{T}^{(0)}}{\delta\Phi_A(x)\delta\Phi_B(y)}\,e^{-ip(x-y)}\right)^{-1} \nnr 
&= \bordermatrix{
& \sml{h^{\gamma\delta}} & \sml{H^{\gamma\delta}} & \sml{b^\gamma} & \sml{c^\gamma} & \sml{\bar{c}^\gamma} \cr
\sml{h^{\alpha\beta}} & \sml{-\dfrac{F^{\alpha\beta\gamma\delta}}{p^2}} & 0 & \sml{-\dfrac{i(\eta^{\alpha\gamma}p^\beta + \eta^{\beta\gamma}p^\alpha)}{p^2}} & 0 & 0 \cr
\sml{H^{\alpha\beta}} &  & \sml{-\dfrac{G^{\alpha\beta\gamma\delta}}{p^2+m^2}} & 0 & 0 & 0 \cr
\sml{b^\alpha} &  &  & 0 & 0 & 0 \cr
\sml{c^\alpha} &  & \hc &  & 0 & \sml{-\dfrac{i\eta^{\alpha\gamma}}{p^2}} \cr
\sml{\bar{c}^\alpha} &  &  &  & & 0}_{\;{\textstyle ,}}
\end{align}
where
\begin{align}
&F^{\alpha\beta\gamma\delta} = \eta^{\alpha\gamma}\eta^{\beta\delta} + \eta^{\alpha\delta}\eta^{\beta\gamma} - \eta^{\alpha\beta}\eta^{\gamma\delta} \\
&G^{\alpha\beta\gamma\delta} = \eta^{\alpha\gamma}\eta^{\beta\delta} + \eta^{\alpha\delta}\eta^{\beta\gamma} - \frac23\eta^{\alpha\beta}\eta^{\gamma\delta} - \frac{1}{m^2}\bigg(\frac23\big(\eta^{\alpha\beta} p^\gamma p^\delta + \eta^{\gamma\delta} p^\alpha p^\beta\big) \nnr
&\phantom{G^{\alpha\beta\gamma\delta} =} - \eta^{\alpha\gamma} p^\beta p^\delta - \eta^{\alpha\delta} p^\beta p^\gamma - \eta^{\beta\delta} p^\alpha p^\gamma - \eta^{\beta\gamma} p^\alpha p^\delta\bigg) + \frac{4}{3m^4}p^\alpha p^\beta p^\gamma p^\delta \,.
\end{align}
One may note the poor UV behavior of the $H$--$H$ propagator in \eqref{propmat} that would seem to imply the theory is non-renormalizable, however, this may be improved by introducing Stückelberg fields and symmetries to recover the $1/p^4$ behavior of the fourth-order propagator, as was done in \cite{Kuntz2022b}. Here we have elected to instead choose the simplest formulation of the second-order theory with the least amount of fields and with only simple-pole propagators in order to keep the discussion as straightforward as possible, since we will not address renormalization in this work.

We proceed by passing to the asymptotic limit \eqref{asymlim} in order to decompose each of our fields in terms of plane wave functions and c/a operators. The EOMs obtained from \eqref{AT0} in the massless boson sector are given by
\begin{align} \label{hbEOMs}
&\mathcal{E}^{(\eta)}_{\alpha\beta\gamma\delta}h^{\gamma\delta} - \partial_{(\alpha} b_{\beta)} + \frac12\mmet\partial_\gamma b^\gamma = 0& &\partial_\beta h_\alpha{}^\beta - \frac{1}{2}\partial_\alpha\hs{\beta} + b_\alpha = 0
\end{align}
which may be combined to find that the graviton obeys the massless Klein-Gordon equation
\begin{align}
\Box\,\hab = 0 \,.
\end{align}
The massless FP ghosts are similarly found to obey the EOMs
\begin{align}
\Box\,c^\alpha = 0 \eqspace \Box\,\bar{c}_\alpha = 0 \,.
\end{align}
In the massive sector, which contains only $\Hab$, we have the familiar Fierz-Pauli EOM
\begin{align}
\mathcal{E}^{(\eta)}_{\alpha\beta\gamma\delta}H^{\gamma\delta} - \frac{m^2}{2}\Big(\Hab - \mmet\Hs{\gamma}\Big) = 0 \,,
\end{align}
whose trace and divergence can be combined to find the restrictions
\begin{align}
\partial_\beta H_\alpha{}^\beta = 0 \eqspace \Hs{\alpha} = 0 \,,
\end{align}
which indicate that $\Hab$ must satisfy the massive Klein-Gordon equation
\begin{align}
\big(\Box - m^2\big)\Hab = 0 \,.
\end{align}
Since all of our fields correspond to simple-pole propagators, it is straightforward to show that the EOMs above are solved by the decompositions
\begin{align} \label{decomps}
\begin{aligned}
&\hab(x) = \hat{h}_{\alpha\beta}(\bm p)f_{\bm p}(x;0) + \hc\qquad\qquad& &\Hab(x) = \hat{H}_{\alpha\beta}(\bm p)f_{\bm p}(x;m) + \hc \\
&b_\alpha(x) = \hat{b}_{\alpha}(\bm p)f_{\bm p}(x;0) + \hc \\
&c^\alpha(x) = \hat{c}^\alpha(\bm p)f_{\bm p}(x;0) + \hc& &\bar{c}_\alpha(x) = \hat{\bar{c}}_\alpha(\bm p)f_{\bm p}(x;0) + \hc \,,
\end{aligned}
\end{align}
paired with the restrictions that result from our chosen gauge condition \eqref{gaugecon} and the gauge/BRST-invariant nature of $\Hab$:
\begin{align} \label{hHconds}
&p^\beta\hat{h}_{\alpha\beta} = \frac12p_\alpha\hat{h}_\beta{}^\beta+ i\hat{b}_\alpha& &p^\beta\hat{H}_{\alpha\beta} = 0& &\hat{H}_\beta{}^\beta = 0 \,.
\end{align}
We recall that $f_{\bm p}$ is a plane-wave function that satisfies $(\Box-m^2)f_{\bm p}(x;m)=0$ and note that we have suppressed sums over $\bm p$ in \eqref{decomps} to avoid clutter.

Finally, the propagator matrix \eqref{propmat} allows us to read off the following non-zero commutators between the fields,
\begin{align}
\begin{aligned}
&\tCom{h_{\alpha\beta}(x)}{h_{\gamma\delta}(y)} = F_{\alpha\beta\gamma\delta}D(x - y;0) \qquad\quad \tCom{h_{\alpha\beta}(x)}{b_\gamma(y)} = -\big(\eta_{\alpha\gamma}\partial_\beta + \eta_{\beta\gamma}\partial_\alpha\big)D(x - y;0) \\
&\tCom{H_{\alpha\beta}(x)}{H_{\gamma\delta}(y)} = \Big(G_{\alpha\beta\gamma\delta}\big|_{p\,=\,i\partial}\Big)D(x - y;m) \qquad\quad \tACom{c_\alpha(x)}{\bar{c}_\beta(y)} = i\mmet D(x - y;0) \,,
\end{aligned}
\end{align}
where the definition of the invariant delta functions $D(x-y;m)$ is given in \eqref{phicom}. These expressions in turn allow us to derive commutators between the state operators after inserting the decompositions \eqref{decomps}:
\begin{align} \label{opcoms}
\begin{aligned}
&\tCom{\hat{h}_{\alpha\beta}(\bm p)}{\hat{h}_{\gamma\delta}(\bm q)} = F_{\alpha\beta\gamma\delta}\delta^3(\bm p - \bm q) \qquad&
&\tCom{\hat{h}_{\alpha\beta}(\bm p)}{\hat{b}_\gamma(\bm q)} = -i\big(\eta_{\alpha\gamma}p_\beta + \eta_{\beta\gamma}p_\alpha\big)\delta^3(\bm p - \bm q) \\
&\tCom{\hat{H}_{\alpha\beta}(\bm p)}{\hat{H}_{\gamma\delta}(\bm q)} = G_{\alpha\beta\gamma\delta}\delta^3(\bm p - \bm q)&
&\tACom{\hat{c}_\alpha(\bm p)}{\hat{\bar{c}}_\beta(\bm q)} = i\mmet\delta^3(\bm p - \bm q) \,.
\end{aligned}
\end{align}
With this, our complete Hilbert space $\FS$, which is spanned by the states created by the operators $\{\hat{h}^\dag_{\alpha\beta},$ \!$\hat{H}^\dag_{\alpha\beta},$ \!$\hat{b}^\dag_\alpha,$ \!$\hat{c}^\dag_\alpha,$ \!$\hat{\bar{c}}^\dag_\alpha\}$, is established and all that remains is an identification of its positive-definite subspace through the Kugo-Ojima quartet mechanism.

\subsection{Transverse physical subspace}

As discussed in Section \ref{subsec:physsub}, the transverse physical subspace of interest is identified based on the BRST transformation of each state in $\FS$. Since we will make these identifications with respect to the quantum c/a operators that define these states, it is convenient to express their BRST transformations in terms of commutators with the BRST charge $\BChg$ as
\begin{align} \label{comtrans}
&\tCom{\BChg}{\hat{h}_{\alpha\beta}} = p_\alpha\hat{c}_\beta + p_\beta\hat{c}_\alpha& &\tCom{\BChg}{\hat{H}_{\alpha\beta}} = 0& &\tCom{\BChg}{\hat{b}_\alpha} = 0& &\tCom{\BChg}{\hat{c}_\alpha} = 0& &\tCom{\BChg}{\hat{\bar{c}}_\alpha} = \hat{b}_\alpha \,.
\end{align}

Beginning with the massless sector, we may simplify our calculations by selecting the particular reference frame corresponding to motion in the $z$ direction,
\begin{align} \label{z0frame}
p^\alpha = \{\omega_0,0,0,\omega_0\} \,,
\end{align}
where $\omega_0 = |\bp|$. This is done without loss of generalization, as the commutators derived in a particular frame are still valid in general \cite{Kugo1978}. Then, using the transformations \eqref{comtrans}, the independent BRST singlets (BRST-invariant operators that satisfy $\Com{\BChg}{a_i}=0$) are easily identified as
\begin{align} \label{a0ops}
&a_{+} = \frac{1}{2}\Big(\hat{h}_{11} - \hat{h}_{22}\Big)& &a_{\times} = \hat{h}_{12} \,,
\end{align}
which, using \eqref{opcoms}, are found to commute with each other according to
\begin{align} \label{a0coms}
\tCom{a_{i}(\bm p)}{a_{j}^\dag(\bm q)} = \delta_{ij}\delta^3(\bm p - \bm q) \,.
\end{align}
The remaining eight components of $\hat{h}_{\alpha\beta}$ and all the components of the BRST field operators $\hat{b}_\alpha$, $\hat{c}_\alpha$, and $\hat{\bar{c}}_\alpha$ can then be fully classified as either dependent on each other through the first condition of \eqref{hHconds}, or independent members of a Kugo-Ojima quartet with the definitions
\begin{align} \label{quartetdefs}
&\big(\pi^{(0)}_\alpha\big) = \frac{1}{\omega_0}\Array{
	\hat{h}_{03} + \frac12\hat{h}_{33} \\
	\hat{h}_{13} \\
	\hat{h}_{23}\\
	\frac12\hat{h}_{33}}& &\delta^{(1)}_\alpha = \hat{c}_\alpha& &\pi^{(-1)}_\alpha = i\hat{\bar{c}}_\alpha& &\delta^{(0)}_\alpha = -i\hat{b}_\alpha \,.
\end{align}
Naturally, these identifications satisfy the defining BRST transformations of each parent and daughter pair,
\begin{align}
&\tCom{\BChg}{\pi^{(0)}_\alpha} = \delta^{(1)}_\alpha& &\tACom{\BChg}{\delta^{(1)}_\alpha} = 0& &\tACom{\BChg}{\pi^{(-1)}_\alpha} = -\delta^{(0)}_\alpha& &\tCom{\BChg}{\delta^{(0)}_\alpha} = 0 \,,
\end{align}
as well as the required non-zero commutation relations
\begin{align}
&\tCom{\delta^{(0)}_\alpha(\bm p)}{\pi^{(0)\,\dag}_\beta(\bm q)} = \eta_{\alpha\beta}\delta^3(\bm p - \bm q)& &\tACom{\pi^{(-1)}_\alpha(\bm p)}{\delta^{(1)\,\dag}_\beta(\bm q)} = \eta_{\alpha\beta}\delta^3(\bm p - \bm q) \,.
\end{align}
With all of the above, it is easy to demonstrate that the quartet relation \eqref{quartetdef} is satisfied,
\begin{align} \label{QGquartet}
\tensor[_\IPo]{\big\langle\delta^{(0)}_\alpha(\bm p)\big|\pi^{(0)}_\beta(\bm q)\big\rangle}{} = \tensor[_\IPo]{\big\langle\pi^{(-1)}_\alpha(\bm p)\big|\BChg\big|\pi^{(0)}_\beta(\bm q)\big\rangle}{} = \tensor[_\IPo]{\big\langle\pi^{(-1)}_\alpha(\bm p)\big|\delta^{(1)}_\beta(\bm q)\big\rangle}{} = \mmet\delta^3(\bm p - \bm q) \,,
\end{align}
thus confirming that the quartet mechanism works as intended in the massless sector of our theory.

For the massive sector, the most convenient reference frame based on motion in the $z$ direction is given by
\begin{align} \label{zmframe}
p^\alpha = \{\omega_m,0,0,0\} \,.
\end{align}
The task here is in fact much simpler than in the massless sector since all of the components of $\hat{H}_{\alpha\beta}$ are BRST invariant,
\begin{align}
\tCom{\BChg}{\hat{H}_{\alpha\beta}} = 0 \,.
\end{align}
This of course means that no quartet mechanism is realized in this sector, however, the five conditions on $\hat{H}_{\alpha\beta}$ in \eqref{hHconds} allow us to eliminate five of its ten components and identify the remaining five as the BRST singlets
\begin{align} \label{amops}
\begin{aligned}
&b_{+} = \frac{1}{2}\Big(\hat{H}_{11} - \hat{H}_{22}\Big)\qquad& &b_{\times} = \hat{H}_{12} \\
&b_{1}  = \hat{H}_{13}\qquad& &b_{2}  = \hat{H}_{23}\qquad& &b_{3}  = \frac{\sqrt{3}}{2}\hat{H}_{33} \,,
\end{aligned}
\end{align}
which commute with each other according to 
\begin{align} \label{amcoms}
\tCom{b_{i}(\bm p)}{b_{j}^\dag(\bm q)} = \delta_{ij}\delta^3(\bm p - \bm q) \,.
\end{align}

This classification of all the independent state operators in our theory as either BRST singlets or members of a quartet fully defines the transverse physical positive-definite subspace $\bFSp$ as the space spanned by states created by the seven $a_{i}^\dag$ and $b_{j}^\dag$ operators in \eqref{a0ops} and \eqref{amops}. This definition is reinforced by the positive commutators \eqref{a0coms} and \eqref{amcoms}, while the derivation of the quartet-defining relation \eqref{QGquartet} solidifies the $\pi^{(0)}$, $\delta^{(1)}$, $\pi^{(-1)}$, and $\delta^{(0)}$ states as members of Kugo-Ojima quartets whose zero norm contributions to $\FSp$ do not threaten the unitarity of the S-matrix on $\bFSp$.

It is now a simple manner to derive the free quantum Hamiltonian operator by noting that $\hat{h}_{\alpha\beta}$ and $\hat{H}_{\alpha\beta}$ may be written as
\begin{align} \label{physdecomp}
&\hat{h}_{\alpha\beta}(\bm p) = \sum_{i=+,\times}\Big(\varepsilon^{(i)}_{\alpha\beta}(\bm p)a_{i} (\bm p)\Big) + \cdots \qquad\qquad \hat{H}_{\alpha\beta}(\bm p) = \sum_{j=+,\cdots,3}\Big(\varepsilon^{(j)}_{\alpha\beta}(\bm p)b_{j} (\bm p)\Big) + \cdots \,,
\end{align}
where the ``$\cdots$'' represent the unphysical components confined to the quartets and the $\varepsilon^{(i)}_{\alpha\beta}$ are orthonormalized spin-2 polarization tensors satisfying $p^\alpha\varepsilon^{(i)}_{\alpha\beta}=0$ and $\varepsilon^{(i)}_{\alpha\beta}\,\varepsilon^{(j)\,\alpha\beta}=2\delta_{ij}$, with the component values
\begingroup
\setlength{\arraycolsep}{5pt}
\renewcommand{\arraystretch}{.8}
\begin{align}
\begin{aligned}
&\big(\varepsilon^{(+)}_{\alpha\beta}\big) =
\begin{pmatrix}
0 & 0 & 0 & 0 \\
0 & 1 & 0 & 0 \\
0 & 0 & -1 & 0 \\
0 & 0 & 0 & 0
\end{pmatrix}&
&\big(\varepsilon^{(\times)}_{\alpha\beta}\big) =
\begin{pmatrix}
0 & 0 & 0 & 0 \\
0 & 0 & 1 & 0 \\
0 & 1 & 0 & 0 \\
0 & 0 & 0 & 0
\end{pmatrix} \\[0.8em]
&\big(\varepsilon^{(1)}_{\alpha\beta}\big) =
\begin{pmatrix}
0 & 0 & 0 & 0 \\
0 & 0 & 0 & 1 \\
0 & 0 & 0 & 0 \\
0 & 1 & 0 & 0
\end{pmatrix}&
&\big(\varepsilon^{(2)}_{\alpha\beta}\big) =
\begin{pmatrix}
0 & 0 & 0 & 0 \\
0 & 0 & 0 & 0 \\
0 & 0 & 0 & 1 \\
0 & 0 & 1 & 0
\end{pmatrix}&
&\big(\varepsilon^{(3)}_{\alpha\beta}\big) =
\frac{1}{\sqrt{3}}\begin{pmatrix}
0 & 0 & 0 & 0 \\
0 & -1 & 0 & 0 \\
0 & 0 & -1 & 0 \\
0 & 0 & 0 & 2
\end{pmatrix}
\end{aligned}
\end{align}
\endgroup
for motion in the $z$ direction as in \eqref{z0frame} and \eqref{zmframe}.

With this, the free transverse physical Hamiltonian may be derived from the Heisenberg equation,
\begin{align}
\Com{\Ham_0}{\phi_a(x)} = -i\partial_0\phi_a(x) \,,
\end{align}
where $\phi_a=\{\hab,\Hab\}$, after applying the decompositions \eqref{physdecomp} and using that $\partial_0f_{\bm p}(x) = -ip^0f_{\bm p}(x)$:
\begin{align} \label{HQG0}
\Ham_0 = \int\dd^3{\bm p}\!\!\!\sum_{\sumoverset{i=+,\times}{j=+,\cdots,3}}\!\!\!\Big(\omega_0\,a_{i}^\dag(\bm p)a_{i}(\bm p) + \omega_m\,b_{j}^\dag(\bm p)b_{j}(\bm p)\Big) \,.
\end{align}
This Hamiltonian satisfies the eigenvalue equations
\begin{align}
\Ham_0\ket{a_{i}} = \omega_0\ket{a_{i}} \eqspace \Ham_0\ket{b_{i}} = \omega_m\ket{b_{i}}
\end{align}
and thus possesses seven independent one-particle eigenstates that form a basis on $\bFSp$; these correspond to two transverse massless $\hab$ modes with eigenvalue $\omega_0$ and five transverse massive $\Hab$ modes with eigenvalue $\omega_m$.

Finally, with a clear definition of the asymptotic positive-definite subspace $\bFSp$ and the free Hamiltonian $\Ham_0$, we can establish unitarity of the transverse physical S-matrix based on the conditions we discussed at the beginning of Section \ref{subsec:physsub}. We take asymptotic completeness and pseudo-unitarity of the S-matrix with respect to the $\psIP$ inner product to be satisfied by assumption, while the final condition, positivity of $\psIP$ on $\bFSp$, turns out to be effectively trivial to satisfy in the present theory. We recall that the $\IPo$ operator that defines our inner product is itself defined by the relation $\IPo\Ham-\Ham^\dag\IPo=0$ which, in the asymptotic limit, depends on the free Hamiltonian $\Ham=\Ham_0$. Recalling the positive commutation relations in both the massless \eqref{a0coms} and massive \eqref{amcoms} spin-2 sectors, we may thus establish the required norm-positivity relations
\begin{align}
\tensor[_{\IPo}]{\braket{a_{i}(\bm p)}{a_{j}(\bm q)}}{} = \delta_{ij}\delta^3(\bm p - \bm q) \eqspace \tensor[_{\IPo}]{\braket{b_{i}(\bm p)}{b_{j}(\bm q)}}{} = \delta_{ij}\delta^3(\bm p - \bm q)
\end{align}
with $\IPo=1$. This is of course simply the statement that the Dirac and pseudo-Hermitian inner products are equivalent in the asymptotic limit of our theory, meaning that the unitarity-defining relation \eqref{unireq} follows in the standard way. This comes as a direct consequence of our second-order formulation of the original fourth-order theory that splits the original complex dynamical variable as $\text{Re}[\yab]=\hab$ and $\text{Im}[\yab]=\Hab$, which enabled us to rewrite the action in terms of a \textit{Hermitian free action} plus an infinite tower of non-Hermitian $\Ord[\alpha_g]$ interaction terms that have no bearing on the unitarity of the theory in the asymptotic limit.

\section{Unitary Interactions} \label{sec:UniInt}

Our demonstration of unitarity for the transverse physical S-matrix relies on the assumption that the S-matrix satisfies $\Smat^\psc\Smat=1$, which is itself based on the assumption that $U=e^{-i\Ham t}$ (where $\Ham$ is the full interacting Hamiltonian) generates unitary time evolution as in \eqref{tevo}. The truly remarkable feature of $\PT$-symmetric theory is that such an evolution operator can be consistently defined even if $\Ham$ is non-Hermitian, as a result of the fact that $\PT$ symmetry implies pseudo-Hermiticity and thus the existence of the unitary inner product \eqref{psip}. Of course, it is also important that the interacting Hamiltonian also possesses only real positive eigenvalues, which is guaranteed if the pseudo-Hermiticity operator admits the decomposition $\IPo=\hIPo^\dag\hIPo$ \cite{Mostafazadeh2002a,Mostafazadeh2002b,Mostafazadeh2002c}. Since the S-matrix is connected to the asymptotic limit of the evolution operator that depends on the full Hamiltonian, the proof of unitarity presented in the last section and the physicality of our theory in general thus rests on the existence of such an $\IPo$ for the full interacting theory.

The best known methods for computing $\IPo$ in interacting QFTs rely on perturbative expansion and require one to solve a set of coupled partial differential equations of the field variables at each desired order for each $n$-point interaction. To our knowledge, the only closed form expression for $\IPo$ that has been calculated in a field theoretical setting is for the simple example of $(1,1)$-dimensional $\lambda i\varphi^3$ theory, and only up to $\Ord[\lambda]$, in \cite{Bender2004a,Bender2004b}. Indeed, one of the biggest drawbacks to the pseudo-Hermitian QFT formalism presented here is the inherent difficulty in deriving $\IPo$. 

Given the highly complicated nature of the present theory that contains ($n\to\infty$)-point derivative interactions of tensor fields, we will thus resign to computing $\IPo$ for a truncated version of the theory. We consider the interacting Hamiltonian 
\begin{align} \label{HQGfull}
\Ham = \Ham_0 + \Ham_\text{int}
\end{align}
with the free part
\begin{align}
\Ham_0 = \int\dd^3{\bm p}\big(\omega_0a^\dag a + \omega_mb^\dag b\big) \,,
\end{align}
which corresponds to \eqref{HQG0} after neglecting the $i$ and $j$ spin indices, which is done simply in an effort to make the presentation of our results more clear. Here we have also suppressed $\bp$-dependence, as we will continue to do when such dependence is obvious. 

The interaction terms in \eqref{HQGfull} come from expanding the metric in \eqref{AQG} as $\met\to\mmet+\frac{\alpha_g}{\Mp}(\hab+i\Hab)$ and thus appear schematically as
\begin{align} \label{AQGint}
\Act_\text{QG}^{(\text{int})} \sim \int\dx\bigg[&\alpha_g\bigg(\frac{1}{m}\Big(h(\partial h)^2 + h(\partial H)^2 + i\big(H(\partial h)^2 + H(\partial H)^2\big)\Big) \nnr
&\phantom{\alpha_g\bigg(}+ \frac{1}{m^3}\Big(h(\Box h)^2 + h(\Box H)^2 + i\big(H(\Box h)^2 + H(\Box H)^2\big)\Big)\bigg) + \Ord\big[\alpha_g^2\big]\bigg] \,.
\end{align}
The Heisenberg picture Hamiltonian operator corresponding to these terms may be derived by decomposing the fields per \eqref{decomps} and dropping terms containing quartet c/a operators. At first order, this corresponds to a sum of all the permutations of terms containing three factors of $a$, $a^\dag$, $ib$, and $ib^\dag$. Though there are ten terms in the complete interaction Hamiltonian at $\Ord[\alpha_g]$ (before also taking spin indices into account), for the sake of avoiding clutter\footnote{We note that all the calculations in this section have also been performed taking all interactions up through $\Ord[\alpha_g^2]$ into account and that including them does not effect the essence of any results.}, we will consider the simplified operator
\begin{align} 
\Ham_\text{int} = \alpha_g \Ham_1 \,
\end{align}
where
\begin{align} \label{HQGint}
\Ham_1 = \int\dd^3{\bm p}\Big(\omega_0\big(c_1a^\dag a a + c_2b^\dag a b + \hc \big) + i\omega_m\big(c_3a^\dag a b + c_4b^\dag b b + \hc\big)\Big) \,.
\end{align}
Here, the $c_i$ represent the $\omega_0(\bp)$ and $\omega_m(\bp)$ dependent coefficient of each term. The exact form of each coefficient is not important for our purposes and writing them this way allows us to more easily track the effects of including both real and imaginary interactions in what follows. Though we will silently drop terms $\Ord[\alpha_g^3]$ moving forward, one should keep in mind that we are working in perturbation theory and that the complete theory contains an infinite tower of $n$-point interaction terms proportional to $\alpha_g^{n-2}$.

Recalling that the c/a operators satisfy the commutation relations \eqref{a0coms} and \eqref{amcoms} (here without the factor of $\delta_{ij}$), one can show that the Hamiltonian \eqref{HQGfull} satisfies the eigenvalue equations
\begin{align}
&\Ham\big|A\big\rangle = \Big(\omega_0 - \alpha_g^2\big(2c_1^2\omega_0 - c_3^2\omega_m\big)\Big)\big|A\big\rangle& &\Ham\big|B\big\rangle = \Big(\omega_m - \alpha_g^2\big(c_2^2\omega_0 - 2c_4^2\omega_m\big)\Big)\big|B\big\rangle
\end{align}
with respect to the eigenstates
\begin{align}
&\big|A\big\rangle = \bigg(a^\dag - \alpha_g\big(c_1a^\dag a^\dag + ic_3a^\dag b^\dag\big) - \alpha_g^2\Big(\frac{ic_2c_3\omega_0}{\omega_0 - \omega_m}b^\dag \nnr
&\phantom{\big|A\big\rangle = \bigg(}- c_1^2a^\dag a^\dag a^\dag - \frac{ic_3\big((c_1 + c_2)\omega_0 + 2c_1\omega_m\big)}{\omega_0 + \omega_m}a^\dag a^\dag b^\dag + \frac{c_3(c_3 + c_4)}{2}a^\dag b^\dag b^\dag\Big)\bigg)\big|0\big\rangle \label{a0i}\\
&\big|B\big\rangle = \bigg(b^\dag - \alpha_g\big(c_2a^\dag b^\dag + ic_4b^\dag b^\dag\big) + \alpha_g^2\Big(\frac{ic_2c_3\omega_m}{\omega_0 - \omega_m}a^\dag \nnr
&\phantom{\big|B\big\rangle = \bigg(} + \frac{c_2(c_1 + c_2)}{2}a^\dag a^\dag b^\dag + \frac{ic_2\big(2c_4\omega_0 + (c_3 + c_4)\omega_m\big)}{\omega_0 + \omega_m}a^\dag b^\dag b^\dag - c_4^2b^\dag b^\dag b^\dag\Big)\bigg)\big|0\big\rangle \,. \label{ami}
\end{align}
It is straightforward to confirm that, since we are dealing with a non-Hermitian theory, $\bra{A}$ and $\bra{B}$ are not left eigenstates of $\Ham$, but rather of $\Ham^\dag$. We also see that the Dirac inner product fails to satisfy the positivity requirement of a physical theory, as its inner product metric is imaginary in the off-diagonal:
\begin{align}
\begin{aligned}
&\big\langle A(\bp)\big|A(\bq)\big\rangle = \Big(1 + \alpha_g^2\big(2c_1^2 + c_3^2\big)\Big)\delta^3(\bp - \bq) \\
&\big\langle B(\bp)\big|B(\bq)\big\rangle = \Big(1 + \alpha_g^2\big(c_2^2 + 2c_4^2\big)\Big)\delta^3(\bp - \bq) \\
&\big\langle A(\bp)\big|B(\bq)\big\rangle = \frac{2i\alpha_g^2 c_2c_3\omega_m}{\omega_0 - \omega_m}\delta^3(\bp - \bq) \,.
\end{aligned}
\end{align}

Despite this unattractive feature, the physicality of our model can be established by appealing to its $\PT$ symmetry/pseudo-Hermiticity. To calculate an explicit expression of the $\IPo$ corresponding to $\Ham$, which will allow us to investigate the structure of the modified inner product metric, we make the ansatz
\begin{align} \label{etaans}
&\IPo = e^{-\eIPo} = 1 - \eIPo + \frac12 \eIPo^2 +\cdots \qquad\text{with}\qquad \eIPo = \alpha_g\eIPo_1 + \alpha_g^2\eIPo_2 + \cdots \,.
\end{align}
The defining relation of the pseudo-Hermiticity operator, $\IPo\Ham-\Ham^\dag\IPo=0$, then takes the form
\begin{align} \label{sigmaeqs}
\alpha_g\Big(\Com{\Ham_0}{\eIPo_1} + \Ham_1 - \Ham_1^\dag\Big) + \alpha_g^2\bigg(\Com{\Ham_0}{\eIPo_2} - \frac12\Com{\Ham_0}{\eIPo_1^2} - \eIPo_1\Ham_1 + \Ham_1^\dag\eIPo_1\bigg) + \cdots = 0 \,,
\end{align}
thus allowing us to solve for the $\eIPo_i$ iteratively at each order in $\alpha_g$. 

The $\eIPo_i$ may generally describe non-local interactions of the c/a operators in phase space, so for the first order part of $\eIPo$, we take an ansatz based on the imaginary part of the interaction Hamiltonian:
\begin{align}
\eIPo_1 = \int\dd^3\bq_1\dd^3\bq_2\dd^3\bq_3\Big(s^{(1,1)}_{\bq_1\bq_2\bq_3}a^\dag_{\bq_1}a_{\bq_2}b_{\bq_3} + s^{(1,2)}_{\bq_1,\bq_2,\bq_3}b^\dag_{\bq_1}b_{\bq_2}b_{\bq_3} +\hc\Big) \,,
\end{align}
where we have indicated functional dependence with subscripts to avoid clutter. With this, we may solve for the arbitrary functions $s^{(1,i)}_{\bq_1\bq_2\bq_3}\equiv s^{(1,i)}(\bq_1,\bq_2,\bq_3)$ by inserting this ansatz into the $\Ord[\alpha_g]$ part of \eqref{sigmaeqs}. After commuting c/a operators so as to arrive at a normal-ordered expression, we find
\begin{align}
 \int\dd^3\bp\dd^3\bq_1\dd^3\bq_2\dd^3\bq_3\Big(&s^{(1,1)}_{\bq_1\bq_2\bq_3}a^\dag_{\bq_1}a_{\bq_2}b_{\bq_3} - 2ic_{3\bp}a^\dag_{\bp}a_{\bp}b_{\bp} \nnr
 &+ s^{(1,2)}_{\bq_1,\bq_2,\bq_3}b^\dag_{\bq_1}b_{\bq_2}b_{\bq_3} - 2ic_{4\bp}b^\dag_{\bp}b_{\bp}b_{\bp} + \hc\Big) = 0 \,,
\end{align}
thus indicating that the solutions
\begin{align}
s^{(1,1)}_{\bq_1\bq_2\bq_3} = 2ic_{3}\delta^3(\bq_1 - \bq_2)\delta^3(\bq_2 - \bq_3) \qquad\qquad s^{(1,2)}_{\bq_1\bq_2\bq_3} = 2ic_{4}\delta^3(\bq_1 - \bq_2)\delta^3(\bq_2 - \bq_3)
\end{align}
yield the first-order contribution to $\eIPo$,
\begin{align}
\eIPo_1 = \int\dd^3\bq\Big(2i\big(c_3a^\dag a b + c_4b^\dag b b\big) + \hc\Big) \,.
\end{align}

To solve for the second order contribution, we insert our expression for $\eIPo_1$ into the $\Ord[\alpha_g^2]$ part of \eqref{sigmaeqs} and make another ansatz based on the possible imaginary combinations of $a$ and $ib$:
\begin{align}
\eIPo_2 = \int\dd^3\bq_1\dd^3\bq_2\dd^3\bq_3\dd^3\bq_4\Big(&s^{(2,1)}_{\bq_1\bq_2}a^\dag_{\bq_1}b_{\bq_2} + s^{(2,2)}_{\bq_1\bq_2\bq_3\bq_4}a^\dag_{\bq_1}a_{\bq_2}a_{\bq_3}b_{\bq_4} + s^{(2,3)}_{\bq_1\bq_2\bq_3\bq_4}a^\dag_{\bq_1}a^\dag_{\bq_2}a_{\bq_3}b_{\bq_4} \nnr
&+ s^{(2,4)}_{\bq_1\bq_2\bq_3\bq_4}b^\dag_{\bq_1}a_{\bq_2}b_{\bq_3}b_{\bq_4} + s^{(2,5)}_{\bq_1\bq_2\bq_3\bq_4}a^\dag_{\bq_1}b^\dag_{\bq_2}b_{\bq_3}b_{\bq_4} + \hc\Big) \,.
\end{align}
Following the same process of commuting c/a operators and collecting terms, we find that the second-order functions once again take the form $s^{(2,i)}=f(c_i)\times(\text{deltas})$, and that the complete second order contribution amounts to
\begin{align}
\eIPo_2 = \int\dd^3\bq\bigg[2i\omega_0\bigg(&\frac{c_2c_3}{\omega_0 - \omega_m}a^\dag b + \frac{(c_1 - c_2)c_3}{\omega_0 + \omega_m}a^\dag a a b + \frac{(c_1 + c_2)c_3}{\omega_0 - \omega_m}a^\dag a^\dag a b \nnr
&+ \frac{c_2(c_3 - c_4)}{\omega_0 + \omega_m}b^\dag a b b + \frac{c_2(c_3 + c_4)}{\omega_0 - \omega_m}a^\dag b^\dag b b\bigg) + \hc\bigg] \,.
\end{align}
One may of course continue this process to arbitrarily higher orders, though the present second-order calculation is enough to see how the adoption of the pseudo-Hermitian paradigm is able to establish the physicality of the present theory. Indeed, the positive-definite-ness of the pseudo-Hermitian inner product defined with respect to $\IPo$ manifests with the norms
\begin{align} \label{psnorms}
\begin{aligned}
&\tensor[_\IPo]{\big\langle A(\bp)\big|A(\bq)\big\rangle}{} = \big\langle A(\bp)\big|\,e^{-\eIPo}\,\big|A(\bq)\big\rangle= \Big(1 + \alpha_g^2\big(2c_1^2 - c_3^2\big)\Big)\delta^3(\bp - \bq) \\
&\tensor[_\IPo]{\big\langle B(\bp)\big|B(\bq)\big\rangle}{} = \big\langle B(\bp)\big|\,e^{-\eIPo}\,\big|B(\bq)\big\rangle = \Big(1 + \alpha_g^2\big(c_2^2 - 2 c_4^2\big)\Big)\delta^3(\bp - \bq) \\
&\tensor[_\IPo]{\big\langle A(\bp)\big|B(\bq)\big\rangle}{} = \big\langle A(\bp)\big|\,e^{-\eIPo}\,\big|B(\bq)\big\rangle = 0 \,,
\end{aligned}
\end{align}
which are diagonal and strictly positive for perturbative $\alpha_g$.

The definition \eqref{etaans} also defines the operator $\hIPo = e^{-\eIPo/2}$ which, as discussed in Section \ref{subsec:gencons}, may be used as a similarity transformation that maps the non-Hermitian Hamiltonian \eqref{HQGfull} to its Hermitian counterpart $\bar{\Ham}=\bar{\Ham}^\dag$:
\begin{align} \label{HQGbar}
\bar{\Ham} \equiv e^{-\eIPo/2} \Ham e^{\eIPo/2} = \bar{\Ham}_0 + \bar{\Ham}_\text{int}
\end{align}
where
\begin{align}
&\bar{\Ham}_0 = \Ham_0 = \int\dd^3{\bm p}\big(\omega_0a^\dag a + \omega_mb^\dag b\big) \\
&\bar{\Ham}_\text{int} = \int\dd^3{\bm p}\bigg(\alpha_g\omega_0\big(c_1a^\dag a a + c_2b^\dag a b + \hc\big) \nnr
&\phantom{\bar{\Ham}_\text{int} = \int\dd^3{\bm p}\bigg(}+ \alpha_g^2\omega_m\Big(c_3^2\big(a^\dag a + a^\dag a^\dag a a\big) + 4c_3c_4a^\dag b^\dag a b + c_4^2\big(2b^\dag b + 3b^\dag b^\dag b b\big)\Big)\bigg) \,.
\end{align}
It is interesting to note that, while the Hermitian $c_1$ and $c_2$ terms in \eqref{HQGint} remain after this transformation, the anti-Hermitian $c_3$ and $c_4$ terms are removed from the $\Ord[\alpha_g]$ part of $\bar{\Ham}_\text{int}$ entirely. The influence of these terms still manifests at the next order however, where they come to parameterize additional Hermitian interactions with even numbers of $a$ and $b$. Though we have not displayed all possible interaction terms here, having performed all the calculations above for the most general set of interactions up through $\Ord[\alpha_g^2]$, we can confirm that this behavior i.e.\ the transfer and transformation of anti-Hermitian interactions to the next order, appears to be a general feature. Interestingly, this can lead to fundamentally different interactions in each form of the theory. For example, the $c_3$ term in \eqref{HQGint} describes $H \to hh$ decay, and while this term still has an effect on the Hermitian theory's interactions at $\Ord[\alpha_g^2]$, it instead parameterizes four-point interactions and such an $H \to hh$ decay is no longer possible.

Finally, it is also important to point out that, as anticipated, the eigenvalues of the original non-Hermitian Hamiltonian are preserved after similarity transformation:
\begin{align}
&\bar{\Ham}\ket{\bar{A}} =  \Big(\omega_0 - \alpha_g^2\big(2c_1^2\omega_0 - c_3^2\omega_m\big)\Big)\ket{\bar{A}}\qquad& &\bar{\Ham}\ket{\bar{B}} = \Big(\omega_m - \alpha_g^2\big(c_2^2\omega_0 - 2c_4^2\omega_m\big)\Big)\ket{\bar{B}} \,,
\end{align}
where the transformed eigenstates are given by
\begin{align}
&\ket{\bar{A}} \equiv e^{-\eIPo/2}\big|A\big> = \bigg(a^\dag - \alpha_gc_1a^\dag a^\dag - \frac{\alpha_g^2}{2}\big(c_3^2a^\dag - 2c_1^2 a^\dag a^\dag a^\dag\big)\bigg)\big|0\big\rangle \\
&\ket{\bar{B}} \equiv e^{-\eIPo/2}\big|B\big> = \bigg(b^\dag - \alpha_gc_2a^\dag b^\dag - \frac{\alpha_g^2}{2}\big(2c_4^2b^\dag + c_2(c_1+c_2) a^\dag a^\dag b^\dag\big)\bigg)\big|0\big\rangle \,.
\end{align}
We also find that the Hermitian inner product metric in this transformed basis is equivalent to the $\PT$ inner product metric in the original basis \eqref{psnorms}, as expected:
\begin{align}
\begin{aligned}
&\big\langle\bar{A}(\bp)\big|\bar{A}(\bq)\big\rangle = \tensor[_\IPo]{\big\langle A(\bp)\big|A(\bq)\big\rangle}{} = \Big(1 + \alpha_g^2\big(2c_1^2 - c_3^2\big)\Big)\delta^3(\bp - \bq) \\
&\big\langle\bar{B}(\bp)\big|\bar{B}(\bq)\big\rangle = \tensor[_\IPo]{\big\langle B(\bp)\big|B(\bq)\big\rangle}{} = \Big(1 + \alpha_g^2\big(c_2^2 - 2 c_4^2\big)\Big)\delta^3(\bp - \bq) \\
&\big\langle\bar{A}(\bp)\big|\bar{B}(\bq)\big\rangle = \tensor[_\IPo]{\big\langle A(\bp)\big|B(\bq)\big\rangle}{} = 0 \,.
\end{aligned}
\end{align}

In summary, we have (perturbatively) established that the present interacting model, which is a truncated version of the full gravitational QFT studied in the previous sections, exhibits strictly positive Hamiltonian eigenvalues and a positive-definite inner product metric, despite the fact that it is non-Hermitian. This comes as a result of the existence of the pseudo-Hermiticity operator $\IPo=\hIPo^\dag\hIPo$, which defines the inner product and guarantees unitary time evolution of the system through the relation $U^\psc U=1$. These results give us very good reason to expect that the full theory of interacting quantum quadratic gravity with its tower of spin-dependent ($n\to\infty$)-point interactions also exhibits unitary time evolution and a positive spectrum of Hamiltonian eigenvalues. As previously established, if this is indeed the case, quantum quadratic gravity is demonstrably unitary following the previously discussed prescription that operates in the asymptotic limit.

\section{Conclusion}

In this work we investigated a potential resolution to the ghost problem in fourth-order theories through a complex deformation of the original dynamic variables and the subsequent application of $\PT$-symmetric (pseudo-Hermitian) quantum theory. We described the ghost problem in detail by examining the quintessential quantum mechanical example of the Pais-Uhlenbeck fourth-order oscillator, which has already been shown to be unitarity when treated as a complex-deformed $\PT$-symmetric theory \cite{Bender2008}. The principle purpose of this work was to demonstrate how an analogous prescription can be applied to establish unitarity and a sensible interpretation of probability in quantum quadratic gravity.

Similar studies, based on the Schrödinger picture,  have been performed on the quantum field theoretical version of the PU oscillator and suggested as a solution to the ghost problem in quadratic gravity in the past \cite{Mannheim2018,Mannheim2020a}, however, since quadratic gravity represents a much more complicated gauge theory of dynamical spacetime, particular care must be taken when attempting to establish its unitarity. In this spirit, we adopted the Heisenberg-picture-based covariant operator formalism that has been used to prove unitarity in Yang Mills theories \cite{Kugo1978,Kugo1978a,Kugo1979a,Kugo1979}, and discussed how it should be extended for application in the pseudo-Hermitian framework. We asserted that the Kugo-Ojima quartet mechanism, which describes how unphysical members of the total Hilbert space are confined, is essentially unaffected by the extension to pseudo-Hermitian QFT if the generalized inner product based on the operator $\IPo$ is defined dynamically from the transverse physical Hamiltonian. When this is the case, $\IPo$ is constructed from a sum of products between transverse physical BRST-singlet state operators only and thus commutes with the BRST charge $\BChg$ and all of the unphysical quartet state operators, meaning that the confinement of unphysical particles can occur in the same manner that it does in Hermitian QFT.

Rather than follow the often-used Ostrogradsky prescription for dealing with fourth-order theories, we rewrote the free fourth-order action for quadratic gravity in an equivalent second-order form by introducing an auxiliary tensor field $\Hab$. After allowing for the metric perturbation $\yab$ to be complex, we identified its imaginary part with $\Hab$ and found that the free auxiliary action naturally diagonalizes into the sum of a linearized Einstein-Hilbert action for the real part of the perturbation, $\hab$, and a massive spin-2 Fierz-Pauli action for $\Hab$. While the free part of our second-order diagonalized action was found to be Hermitian, it is accompanied by an infinite tower of non-Hermitian interactions terms. In the standard version of this theory, $\yab$ is assumed to be real and the $\Hab$ part of the action comes with an overall minus sign that identifies it as a massive spin-2 ghost, however, when $\yab$ is taken to be complex as in this work, this troublesome minus sign is eliminated from the free part of the action. This feature is one of the key benefits of the formalism we have presented here -- with a healthy (non-ghostly) and Hermitian free action, demonstration of unitarity and a probability interpretation follow directly from the same BRST methods used to show these features in Yang Mills theories, which operate in the asymptotic limit where only the free theory plays a direct role and the Dirac and pseudo-Hermitian inner products are found to be equivalent.

It is however important to recall that the proof of unitarity based on the free theory relies on the assumption that the S-matrix is pseudo-unitarity, which is only a logical assumption to make if the interacting theory exhibits unitary time evolution. This feature is guaranteed in Hermitian theories since the time evolution operator $U(t)=e^{-i\Ham t}$ is unitary by definition. Though $U$ is not unitary in the usual sense ($U^\dag U=1$) for a general non-Hermitian Hamiltonian, it does satisfy the more general pseudo-unitarity requirement $U^\psc U=1$ if the Hamiltonian is pseudo-Hermitian ($\Ham^\psc=\Ham$) with respect to the pseudo-Hermiticity operator $\IPo$. This operator is what actually defines the positive-semi-definite inner product on the physical Hilbert space $\FSp$ and implies the existence of strictly positive Hamiltonian eigenvalues when it can be written as $\IPo=\hIPo^\dag\hIPo$ \cite{Mostafazadeh2002b}. The existence of such an $\IPo$ in the interacting theory is thus of crucial importance for the proof of unitarity in the asymptotic limit. This link between the interacting and asymptotic theory is not present in Hermitian QFT and it leads to practical difficulties in pseudo-Hermitian QFT since the form of $\IPo$ depends on the Hamiltonian itself through the defining relation $\IPo\Ham-\Ham^\dag\IPo=0$.

In summary, the physicality of non-Hermitian theories relies on the Hamiltonian possessing an unbroken anti-linear ($\PT$) symmetry, or equivalently, on the existence of an operator $\IPo$ that satisfies $\IPo\Ham-\Ham^\dag\IPo=0$ and admits the decomposition $\IPo=\hIPo^\dag\hIPo$. A key result of this work is our identification of the fact that finding a closed form for the pseudo-Hermiticity operator is trivial ($\IPo=1$) in the asymptotic limit when a given pseudo-Hermitian theory possesses a Hermitian free part, as our formulation of QG does. This paired with our assertion that only the transverse physical gauge-fixed Hamiltonian should be used to determine $\IPo$, implies that the proof of unitarity in scattering events follows from the well-established requirements laid out in \cite{Nakanishi1990}, after upgrading the Dirac inner product to ${}_{\IPo}\langle\cdot\rangle$ and assuming that the S-matrix satisfies $\Smat^\psc\Smat=1$. The only caveat to this picture is that this last assumption is based on knowledge of the interacting theory, which is generally very difficult to establish absolutely.

Indeed, as we discussed at the beginning of Section \ref{sec:UniInt}, the only major downside to the pseudo-Hermitian framework lies in the practical difficulty of calculating $\IPo$ in interacting quantum field theory. Due to the highly complicated nature of the interaction terms in quantum quadratic gravity as we have presented it, a complete non-perturbative derivation of $\IPo$ is beyond the scope of this work. However, our perturbative calculation in the truncated model of interacting QG in the previous section does provide compelling evidence that quadratic gravity represents a valid, renormalizable and unitary description of quantum gravity.

Though one may naturally wonder if a theory whose inner product is so difficult to explicitly define is of any practical use, there is a silver lining to this picture. As briefly discussed at the end of Section \ref{subsec:gencons}, correlators in QFT may be calculated essentially as normal in the path integral formalism, without specific knowledge of the $\IPo$ operator. This comes as a result of the fact that the mapping from the original theory to its Hermitian counterpart is also encoded in the couplings between observables and the external current \cite{Jones2007}. This convenient feature has been taken advantage of in the explicit calculation of renormalized correlators in $\lambda\phi^2(i\phi)^\varepsilon$ theory in the past, and is expected to hold as a general statement in pseudo-Hermitian QFT \cite{Bender2018a,Felski2021a}. This puts the ease of practical application of non-Hermitian QG on par with its Hermitian counterpart and relegates the most difficult calculations that require a closed form $\IPo$ to the regime of formal applications in operator-based QFT.

Finally, we note a few interesting avenues for future research that are opened up by the present work. First among these is an investigation of the role of radiative corrections in the present theory, with particular emphasis on the self-energy of the would-be-ghost $\Hab$. As previously mentioned, radiative corrections generate a complex mass for $\Hab$ in the Hermitian formulation of QG (where it is a ghost), and though it is claimed that higher derivative pseudo-Hermitian $\phi^4$ theory is free of complex masses, it is not yet clear if the same is true in pseudo-Hermitian QG due to its more complicated structure of interactions. Describing loop-order effects in the present theory is of course crucial if we are to gain an understanding of its phenomenology. It would also be very interesting to apply the formalism presented here to fourth-order actions describing the fundamental forces of the Standard Model. As mentioned at the end of Section \ref{subsec:compdefQG}, the issues with the original Lee-Wick formulation related to its modified $i\epsilon$-prescription are not present in the pseudo-Hermitian formulation of fourth-order theories as we have presented them. It would thus be interesting to try and upgrade the previously studied ``Lee-Wick Standard Model'' (see e.g.\ \cite{Grinstein2008}) to this framework in order to achieve a potential solution to the hierarchy problem in the second-order SM, while avoiding the usual issues present in the Hermitian version of the theory. To our knowledge, this notion of a $\PT$-symmetric Lee-Wick SM has only briefly been discussed in the literature \cite{Shalaby2009,Mannheim2020a}. Optimistically, these ideas could lead to an expanded understanding of not just gravity, but all of the fundamental forces, through their descriptions in terms of ghost-free higher derivative actions.

\tocless\section{Acknowledgments}

We would like to thank Jisuke Kubo and Philip Mannheim for helpful discussions. Many of the calculations in this work were facilitated by the \href{http://www.xact.es}{\texttt{xAct}} suite of packages for \texttt{Wolfram Mathematica}, in particular \cite{Martin-Garcia2008,Brizuela2009,Nutma2014a,Frob2020}.

\bibliography{library}
\bibliographystyle{utphys}

\end{document}